\documentclass[physrev,reprint,showkeys,amsfonts,amssymb,amsmath]{revtex4-1}
\usepackage[T1]{fontenc}
\usepackage[utf8]{inputenc}
\usepackage[spanish,english]{babel}
\usepackage{hyperref}
\hypersetup{hidelinks}
\usepackage{dcolumn}
\usepackage{physics}
\usepackage{enumerate}
\usepackage{slashed}

\newcommand{\MeV}{\text{MeV}}

\newcommand{\fm}{\text{fm}}
\newcommand{\theor}{\text{Theor}}
\newcommand{\expt}{\text{Expt}}

\begin{document}

\title{Radiative decays in charmonium beyond the p/m approximation}
\author{R. Bruschini}
\email{roberto.bruschini@ific.uv.es}
\affiliation{\foreignlanguage{spanish}{Departamento de Física Teórica, IFIC, Universidad de Valencia-CSIC, E-46100 Burjassot (Valencia)}, Spain}
\author{P. González}
\email{pedro.gonzalez@uv.es}
\affiliation{\foreignlanguage{spanish}{Departamento de Física Teórica, IFIC, Universidad de Valencia-CSIC, E-46100 Burjassot (Valencia)}, Spain}


\begin{abstract}
We analyze the theoretical description of radiative decays in charmonium. We use an elementary emission decay model to build the most general electromagnetic transition operator. We show that accurate results for the widths can be obtained from a simple quark potential model reasonably fitting the spectroscopy if the complete form of the operator is used instead of its standard p/m approximation and the experimental masses are properly implemented in the calculation.
\end{abstract}

\maketitle

\section{Introduction\label{SI}}

Electromagnetic decays in heavy quarkonium (bottomonium or charmonium) may play a key role in the understanding of its structure. The current impossibility to directly solve QCD, the theory of strong interactions, for the description of hadrons, forces us to rely, for the knowledge of such structure, on models and/or effective theories, incorporating at the greater extent the properties of QCD (see for instance \cite{Bra04,Bra11} and references therein). Among these approximations the most successful one regarding the number of described heavy quarkonium states below the open flavor meson-meson thresholds is undoubtedly the constituent quark model, see \cite{Bra04} and references therein, where heavy quarkonium is described as a quark-antiquark bound system. Then, as the electromagnetic transition operator is known (with no free parameter) the comparison of the measured widths with their calculated values from different spectroscopic quark models may be an ideal test to discriminate between these models and to advance in the understanding of the heavy quarkonium structure.

In a recent paper \cite{Br3-19}, we have shown that this discrimination may be rather difficult in bottomonium. By using the standard expansion of the electromagnetic transition operator up to $\vb*{p}_{b}/M_{b}$ order, where $\vb*{p}_{b}$ $(M_{b})$ stands for three-momentum (mass) of the $b$ quark, we have shown that accurate results for the widths can be obtained from different quark potential models reasonably fitting the spectroscopy once the experimental masses of the bottomonium states instead of the calculated ones are properly implemented in the calculation. The argument justifying this substitution (and giving meaning to the qualificative ``reasonably fitting the spectroscopy'' employed) is that the experimental masses can be hopefully obtained from the wavefunctions of the calculated states by applying first order perturbation theory. Therefore, the implementation of the experimental masses allows for a direct test of the quark model wavefunctions.

In this article we analyze electromagnetic decays in charmonium, focusing on the quantitatively most relevant electromagnetic transitions, ${{^{3}\!S_1} \leftrightarrow {^{3}\!P_J}}$, for which there are data available. We first show that the $\vb*{p}_{c}/M_{c}$ order approximation, where the subindex $c$ stands for the $c$ quark, does not give rise to such an accurate description of the decay widths as in bottomonium. This could be somehow expected since the expectation value of $\abs{\vb*{p}_{c}} /M_{c}$, representing the speed of the quark, can be about half the speed of light for the low lying charmonium states, what makes the use of the transition operator up to the $\vb*{p}_{c}/M_{c}$ order debatable. We proceed then to build the complete transition operator and to apply it to the calculation of the decay widths. This allows us to to discriminate between different quark models according to their accuracy in the description of radiative decays.

These contents are organized as follows. In Section~\ref{SII} we detail the Cornell potential models we use to calculate the masses and wavefunctions of the low lying charmonium states. We expect that the experimental masses can be obtained from the calculated values via corrections to the Hamiltonian evaluated at first order in perturbation theory. In Section~\ref{SIII} the elementary emission model for electromagnetic transitions is developed in some detail. From it we recover the usual $\vb*{p}_{c}/M_{c}$ approximation in Section~\ref{SIV}. The comparison of the calculated decay widths with data points out the need to go beyond this approximation. Then, in Section~\ref{SV} the complete transition operator, to all $\vb*{p}_{c}/M_{c}$ orders, is built. The transition amplitude and the explicit form of its electric and magnetic contributions are detailed in Section~\ref{SV} and applied in Section~\ref{SVI} to the calculation of radiative decay widths which are compared to data. Finally, in Section~\ref{SVII} our main results and conclusions are summarized.

\section{Spectroscopic Quark Models\label{SII}}

For the description of charmonium we shall use a nonrelativistic quark potential model framework defined by the Hamiltonian
\begin{equation}
H=\frac{\vb*{p}^{2}}{M_{c}}+V(r),
\label{ham}
\end{equation}
with a Cornell potential energy
\begin{equation}
V(r)  =\sigma r-\frac{\zeta}{r}+\beta
\label{pot}
\end{equation}
where $\vb*{p}$ is the relative momentum operator, $r=\abs{\vb*{r}}$ is the $c-\overline{c}$ distance operator ($\vb*{r}$ is the relative position operator), the parameters $\sigma$ and $\zeta$ stand for the string tension and the chromoelectric coulomb strength respectively, and $\beta$ is a constant to fix the origin of the potential. It is important to emphasize that
\begin{enumerate}[i)]
\item this potential form arises from spin
independent quenched lattice QCD calculations in the Born-Oppenheimer approximation \cite{Bal01},
\item in the spirit of the nonrelativistic quark
model calculations $\sigma$, $\zeta$, $\beta$ and the quark mass $M_{c}$ should be considered as effective parameters through which spin dependent and/or spin independent corrections may be implicitly incorporated.
\end{enumerate}

Henceforth we shall make use of two different quark models with the same Hamiltonian form \eqref{ham}, Model~I and Model~II, that have been used for the analysis of radiative decays in bottomonium \cite{Br3-19}. As we are dealing with a radial potential we shall denote the spectroscopic states by $n\,{^{2s+1}\!L_{J}}$ where $s,$ $L$ and $J$ stand for the spin, orbital angular momentum and total angular momentum quantum numbers respectively.

Model~I, providing a good description of spin triplet state masses in bottomonium \cite{Gon14} is defined in charmonium \cite{Gon15} by the set of parameter values
\begin{equation}
\begin{split}
\sigma_{I}&=850\,\MeV/\fm,\\
\zeta_{I} &=100 \, \MeV \cdot \fm, \\
(M_{c})_{I}&=1348.6 \, \MeV,
\end{split}
\label{par1}
\end{equation}
giving account of the mass differences between some of the low lying (spin triplet) charmonium states whose electromagnetic transitions are measured, more precisely between $2{^{3}\!S_{1}}$ and $1{^{3}\!P_{1}}$, and between $1{^{3}\!P_{1}}$ and $1{^{3}\!S_{1}}$. Hence, the model also describes accurately, through a convenient choice of the additive constant $(\beta_{c})  _{I}$, the masses of the $1{^{3}\!S_{1}}$, $2{^{3}\!S_{1}}$ and $1{^{3}\!P_{1}}$ states. Furthermore, inasmuch as the mass splittings between $P$ states can be obtained via first order perturbation theory the $1{^{3}\!P_{0,2}}$ states would be described by the same wavefunction as the $1{^{3}\!P_{1}}$ one. In Table~\ref{tabmassdif1} we list the ratios of the calculated mass differences to the experimental ones.
\begin{table}
\begin{ruledtabular}
\begin{tabular}{ld}
$i-f$ & \multicolumn{1}{c}{$\frac{(M_{i}-M_{f})_I^\theor}{(M_{i}-M_{f})^\expt}$}\\
\hline
$2{^{3}\!S_{1}}-1{^{3}\!P_{1}}$ & 1.00\\
$1{^{3}\!P_{1}}-1{^{3}\!S_{1}}$ & 0.99\\
$2{^{3}\!S_{1}}-1{^{3}\!S_{1}}$ & 0.99\\
$2{^{3}\!S_{1}}-1{^{3}\!P_{2}}$ & 1.09\\
$2{^{3}\!S_{1}}-1{^{3}\!P_{0}}$ & 0.65\\
$1{^{3}\!P_{0}}-1{^{3}\!S_{1}}$ & 1.29\\
$1{^{3}\!P_{2}}-1{^{3}\!S_{1}}$ & 0.89\\
\end{tabular}
\end{ruledtabular}
\caption{Ratios of the calculated mass differences $(M_{i}-M_{f})_I^\theor$ between $1{^{3}\!S_{1}},$ $2{^{3}\!S_{1}}$, $1{^{3}\!P_{1}}$ charmonium states from Model~I as compared to the experimental ones $(  M_{i}-M_{f})^\expt$ taken from \cite{PDG18}.}
\label{tabmassdif1}
\end{table}

Model~II, defined in reference \cite{Eic94} by the set of parameter values
\begin{equation}
\begin{split}
\sigma_{II}&=925.5\,\MeV/\fm, \\
\zeta_{II}&=102.6 \, \MeV \vdot \fm, \\
(M_{c})_{II}&=1840 \, \MeV,
\end{split}
\label{par}
\end{equation}
is based on the assumption that mass corrections to the charmonium states calculated from \eqref{ham} may have to do mainly with non considered spin dependent terms in the potential, so that the quark model should fit the centers of gravity of spin triplet and spin singlet states.

Let us realize that the chosen value for the string tension, $\sqrt{\sigma_{II}} = 427.4$ MeV, agrees with the one derived from the analysis of Regge trajectories in light mesons \cite{Bal01}, and that the Coulomb strength $\zeta_{II}=102.6 \MeV \fm$ corresponds to a strong quark-gluon coupling $\alpha_{s}=\frac{3\zeta}{4\hbar}\simeq 0.39$ quite in agreement with the value derived in QCD from the fine structure splitting of $1P$ states in charmonium \cite{Bad99}. As for $M_{c}$ the chosen value gives mass differences between any two of the centers of gravity of the $1S$, $2S$ and $1P$ states in accord with data within a $10\%$ of accuracy. In Table~\ref{tabmassdif} we list the ratios of the calculated center of gravity mass differences to the experimental ones. One could think of refining the values of the parameters to make all these ratios closer to $1$. This can only be done at the price of loosing the close connection of $\sigma_{II}$ and/or $\zeta_{II}$ with their expected phenomenological values. Instead, we prefer to maintain such connection and the modest discrepancy reflected in Table~\ref{tabmassdif}. Moreover, the chosen values of $\sigma_{II}$ and $\zeta_{II}$ have been used in bottomonium for an accurate description of electromagnetic decay widths \cite{Br3-19} what adds interest to the possibility of getting such an accurate description in charmonium as well.

\begin{table}
\begin{ruledtabular}
\begin{tabular}{ld}
$i-f$ & \multicolumn{1}{c}{$\frac{(\overline{M}_{i}-\overline{M}_{f})_{II}^\theor}{(\overline{M}_{i}-\overline{M}_{f})^\expt}$}\\
\hline
$2S-1S$ & 0.98\\
$2S-1P$ & 1.09\\
$1P-1S$ & 0.94\\
\end{tabular}
\end{ruledtabular}
\caption{Ratios of the calculated center of gravity mass differences $(\overline{M}_{i}-\overline{M}_{f})_{II}^\theor$ between $1S$, $2S$, $1P$ charmonium states from Model~II as compared to the experimental ones $(\overline{M}_{i}-\overline{M}_{f})^\expt$ taken from \cite{PDG18}.}
\label{tabmassdif}
\end{table}

It is worth to point out that the quark model wavefunctions are the same for any value of the additive constant $\beta$. Indeed, we could change $(\beta_{c})_{II}$ to obtain from Model~II an approximate mass description of the $1{^{3}\!S_{1}},$ $2{^{3}\!S_{1}}$ and $1{^{3}\!P_{1}}$ states, or we could change $(\beta_{c})_{I}$ to obtain from Model~I the centers of gravity of the $1S,$ $2S$ and $1P$ states in accord with data within a $5\%$ of accuracy. In this sense, Model~I and Model~II are quite equivalent with respect to the mass description of the low lying charmonium states except for $2{^{3}\!P_J}$, as shown in Table~\ref{tabIII}. Hence, any significant difference in their predictions for other observables involving these states can put severe constraints on the possible values of the parameters, in particular on the charm mass whose value in Model~I is very different to that in Model~II.

\begin{table}
\begin{ruledtabular}
\begin{tabular}{lccc}
$n\,{^{2s+1}\!L_{J}}$ & $(M_{n\,{^{2s+1}\!L_{J}}})_{I}$ & $(M_{n\,{^{2s+1}\!L_{J}}})_\expt$ & $(M_{n\,{^{2s+1}\!L_{J}}})_{II}$\\
& MeV & MeV & MeV \\
\hline
$1{^{3}\!S_{1}}$ & $3099$ & $3096.916\pm0.011$ & $3088$\\
$2{^{3}\!S_{1}}$ & $3685$ & $3686.09\pm0.04$ & $3678$\\
$1{^{3}\!P_{0}}$ & $3509$ & $3414.75\pm0.31$ & $3516$\\
$1{^{3}\!P_{1}}$ & $3509$ & $3510.66\pm0.07$ & $3516$\\
$1{^{3}\!P_{2}}$ & $3509$ & $3556.20\pm0.09$ & $3516$\\
$2{^{3}\!P_{0}}$ & $3911$ & $3862_{-32-13}^{+26+40}$ & $3959$\\
$2{^{3}\!P_{1}}$ & $3911$ &  & $3959$\\
$2{^{3}\!P_{2}}$ & $3911$ & $3927.2\pm2.6$ & $3959$\\
\end{tabular}
\end{ruledtabular}
\caption{Calculated masses, $M_{n\,{^{2s+1}\!L_{J}}}$ for the low lying spin triplet states. The subscripts $I$ and $II$ refer to Model~I with $(\beta_{c})  _{I}=53 \, \MeV$ and Model~II with $(\beta_{c})_{II}=-850 \, \MeV$ respectively. Experimental masses from \cite{PDG18}. We have not quoted any mass for the experimental $2{^{3}\!P_{1}}$ state since the $X(3872)$ can not be considered a pure Cornell state.}
\label{tabIII}
\end{table}

It should be also remarked that simplicity is not the only argument to use a Cornell potential model for the study of radiative decays. The absence of any momentum dependence in the potential allows for a complete factorization of the heavy quarkonium mass dependence in the calculation of the electromagnetic decay widths up to $\vb*{p}_{c}/M_{c}$ order. Furthermore, such a complete factorization can be also pursued to higher orders as will be shown later on. Then, if the experimental masses instead of the calculated ones are implemented in the calculation the comparison of the calculated decay widths to data becomes a powerful tool to test the model wavefunctions. As a counterpart, the use of the Cornell potential in charmonium restricts the study to the low lying states where meson-meson threshold effects can be neglected.

\section{Electromagnetic Decay Model\label{SIII}}

Let us consider the decay $I\rightarrow\gamma F$ where $I$ and $F$ are the initial and final charmonium states respectively. In the rest frame of the decaying meson $I$ the total width is given by (we follow the PDG conventions, see \cite[p.~567]{PDG18})
\begin{equation}
\Gamma_{I\rightarrow\gamma F}=\frac{k_{0}}{8\pi M_{I}^{2}}\frac{1}{(2J_{I}+1)}\sum_{\lambda=\pm1}\sum_{m_{I},m_{F}}\abs{\mathcal{M}_{J_{F},m_{F},J_{I},m_{I}}^{\lambda}}^{2},
\label{width}
\end{equation}
where $k_{0}$ is the energy of the photon and $M_{I}$, $J_{I}$ and $m_{I}$ stand for the mass of $I$, its total angular momentum and its third projection respectively. The polarization of the photon is represented by $\lambda$ (as usual we choose the three-momentum of the photon in the $Z$ direction) and the transition amplitude by $\mathcal{M}_{J_{F},m_{F},J_{I},m_{I}}^{\lambda}$. This amplitude can be obtained from the interaction Hamiltonian $\mathcal{H}^{int}$ as
\begin{multline}
(2\pi)^{3}\delta^{(3)}(\vb*{P}_{I}-\vb*{k}-\vb*{P}_{F})\mathcal{M}_{J_{F},m_{F},J_{I},m_{I}}^{\lambda}=\\
\sqrt{2M_{I}}\sqrt{2E_{F}}\sqrt{2k_{0}}\mel{F\gamma}{\mathcal{H}^{int}}{I},
\label{deltas}
\end{multline}
where $P_{I}=(E_{I},\vb*{P}_{I})  =(M_{I},\vb*{0})$, $P_{F}=(  E_{F},\vb*{P}_{F})$ and $(k_{0},\vb*{k})$ are the meson and photon four-momenta.

In the Elementary Emission Decay Model the radiative transition $I\rightarrow\gamma F$ takes place through the emission of the photon by the quark or the antiquark in the initial state. In QED the interaction Hamiltonian at the quark level is given by
\begin{equation}
\mathcal{H}^{int}=\int \dd{\vb*{x}} j^{\mu}(\vb*{x})  A_{\mu}(\vb*{x}),
\label{Hint}
\end{equation}
where $A_{\mu}$ is the photon field and $j^{\mu}$ the electromagnetic current given by
\begin{equation}
j^{\mu}(\vb*{x})= \overline{q}(\vb*{x}) \mathcal{Q} \gamma^{\mu}q(\vb*{x})
\label{emcurr}
\end{equation}
where $q(\vb*{x})$ is the quark field and $\mathcal{Q}$ the quark charge matrix. Notice that the time dependence has been obviated since in the calculation of the transition amplitude it only gives rise to a Dirac delta accounting for energy conservation.

In quantum field theory the quark field operator with no time dependence is written as (see for example \cite[p.~58]{Pes95}, but note that we use Dirac spinors instead of the Weyl representation adopted there)
\begin{multline}
q(\vb*{x})  =\int\frac{\dd{\vb{p}}}{(2\pi)^{3}}\frac{1}{\sqrt{2\mathrm{E}(\vb{p})}}\\
\sum_{m_{s}}\bigl( u^{m_{s}}(\vb{p})  b_{1}^{m_{s}}(  \vb{p})e^{i\vb{p}\vdot\vb*{x}} + v^{m_{s}}(\vb{p}) b_{2}^{m_{s}\dagger}(\vb{p})e^{-i\vb{p}\vdot\vb*{x}}\bigr),
\label{qfield}
\end{multline}
where $\mathrm{E}(\vb{p})$ is the relativistic energy of a quark $Q$ or antiquark $\bar{Q}$, $\mathrm{E}(\vb{p}) =\sqrt{M_{Q}^{2}+\vb{p}^{2}}$, $u^{m_{s}}(\vb{p})$ and $v^{m_{s}}(\vb{p})$ are the Dirac spinors
\begin{equation}
u^{m_{s}}(\vb{p})  =\sqrt{M_{Q}+\mathrm{E}(\vb{p})}\pmqty{\chi^{m_{s}}\\\frac{\vb{p}\vdot\vb*{\sigma}}{M_{Q}+\mathrm{E}(\vb{p})}\chi^{m_{s}}}
\label{w1}
\end{equation}
\begin{equation}
v^{m_{s}}(\vb{p})  =\sqrt{M_{\bar{Q}}+\mathrm{E}(\vb{p})}\pmqty{\frac{\vb{p}\vdot\vb*{\sigma}}{M_{\bar{Q}}+\mathrm{E}(\vb{p})}\chi^{m_{s}} \\ \chi^{m_{s}}} \label{w2}
\end{equation}
with $\chi^{m_{s}}$ being the Pauli spinor, and $b_{1}^{m_{s}}(\vb{p})$ $(  b_{2}^{m_{s}\dagger}(  \vb{p}))$ the annihilation (creation) operator of a quark (antiquark) with spin projection $m_{s}$ and three-momentum $\vb{p}$.

Then, by defining
\begin{equation}
q_{\alpha}(\vb*{x})  \!=\!\int\frac{\dd{\vb{p}_\alpha}}{(2\pi)^{3}}\sqrt{\frac{M_{\alpha}+\mathrm{E}_\alpha(\vb{p}_\alpha)}{2\mathrm{E}_\alpha(  \vb{p}_\alpha)  }}\sum_{m_{s}}e^{i\vb{p}_\alpha\vdot\vb*{x}} \chi_{\alpha}^{m_{s}}b_{\alpha}^{m_{s}}(\vb{p}_\alpha),
\end{equation}
where $\alpha=1,2$ refers to quark and antiquark respectively, the electromagnetic current $j^{\mu}(\vb*{x})$ reads
\begin{widetext}
\begin{equation}
j^0( \vb*{x}) = \sum_{\alpha=1,2} e_\alpha \Biggr[q_\alpha^\dagger(\vb*{x})q_\alpha(\vb*{x})+ \pqty{\frac{\grad q_\alpha(\vb*{x})^\dagger}{M_\alpha + {E}_\alpha}\vdot\frac{\grad q_\alpha(\vb*{x})}{M_\alpha + {E}_\alpha}+ i \vb*{\sigma}\!_{\alpha}\vdot \frac{\grad q_\alpha(\vb*{x})^\dagger}{M_\alpha + {E}_\alpha} \cp \frac{\grad q_\alpha(\vb*{x})}{M_\alpha + {E}_\alpha}}\Biggr]
\end{equation}
and
\begin{multline}
\vb*{j}(\vb*{x}) = \sum_{\alpha=1,2} e_\alpha \left\{ -i \bqty{q_\alpha^\dagger(\vb*{x}) \pqty{\frac{\grad{q_\alpha(\vb*{x})}}{M_\alpha + {E}_\alpha}}-\pqty{\frac{\grad{q_\alpha^\dagger(\vb*{x})}}{M_\alpha +{E}_\alpha}} q_\alpha(\vb*{x})}+\right.\\
\left.\bqty{\pqty{\frac{\grad{q_\alpha^\dagger(\vb*{x})}}{M_\alpha + {E}_\alpha}} \cp \vb*{\sigma}\!_{\alpha} \, q_\alpha(\vb*{x})-q_\alpha^\dagger(\vb*{x}) \vb*{\sigma}\!_{\alpha} \cp \pqty{\frac{\grad{q_\alpha(\vb*{x})}}{M_\alpha + {E}_\alpha}}}\right\},
\end{multline}
where $e_1 = e_{c}=\frac{2}{3}e$ (with $e=\sqrt{4\pi\alpha_{em}}$ being $\alpha_{em}$ the fine structure constant), $e_2 = e_{\overline{c}} = - e_c$, and $E_{\alpha}=\sqrt{M_{\alpha}^{2}-\vb*{\nabla}_\alpha^{2}}$.

It is important to highlight that in the derivation of the current operator we have kept only terms that conserve separately the number of quarks and antiquarks, since we are only interested in radiative processes where there is no quark-antiquark photoproduction or annihilation.

In order to calculate the matrix element $\mel{F\gamma}{\mathcal{H}^{int}}{I}$ from the spectroscopic quark model wavefunctions one needs the ``first quantized'' form of the interaction. Following the procedure explained in \cite{LeY88} and detailed in Appendix~\ref{firstop}, we obtain the first quantized form of the current (in Appendix~\ref{gaugeinv} it is checked that this current is conserved, as required by gauge invariance)
\begin{multline}
j^0_\text{1st}(\vb*{x}) = \sum_{\alpha=1,2} e_\alpha \sqrt{\frac{M_\alpha + {{E}}_\alpha}{2 {{E}}_\alpha}} \Biggr[\delta^{(3)}(\vb*{x}-\vb*{r}_\alpha)+ \\
\left(\frac{{\vb*{p}}_\alpha}{M_\alpha + {{E}}_\alpha}\vdot \delta^{(3)}(\vb*{x}-\vb*{r}_\alpha) \frac{{\vb*{p}}_\alpha}{M_\alpha + {{E}}_\alpha}+ i \vb*{\sigma}\!_{\alpha}\vdot \frac{{\vb*{p}}_\alpha}{M_\alpha + {{E}}_\alpha} \cp \delta^{(3)}(\vb*{x}-\vb*{r}_\alpha) \frac{{\vb*{p}}_\alpha}{M_\alpha + {{E}}_\alpha}\right)\Biggr] \sqrt{\frac{M_\alpha + {{E}}_\alpha}{2 {{E}}_\alpha}},
\label{cur0}
\end{multline}
\begin{multline}
\vb*{j}_{\text{1st}}(\vb*{x}) = \sum_{\alpha=1,2} e_\alpha \sqrt{\frac{M_\alpha + {{E}}_\alpha}{2 {{E}}_\alpha}} \left[\pqty{\delta^{(3)}(\vb*{x}-\vb*{r}_\alpha) \frac{{\vb*{p}}_\alpha}{M_\alpha + {{E}}_\alpha} + \frac{{\vb*{p}}_\alpha}{M_\alpha + {{E}}_\alpha} \delta^{(3)}(\vb*{x}-\vb*{r}_\alpha)}\right.\\
\left.-i \vb*{\sigma}\!_{\alpha} \cp \pqty{\delta^{(3)}(\vb*{x}-\vb*{r}_\alpha) \frac{{\vb*{p}}_\alpha}{M_\alpha + {{E}}_\alpha} - \frac{{\vb*{p}}_\alpha}{M_\alpha + {{E}}_\alpha} \delta^{(3)}(\vb*{x}-\vb*{r}_\alpha)}\right]\sqrt{\frac{M_\alpha + {{E}}_\alpha}{2 {{E}}_\alpha}}.
\label{curvec}
\end{multline}

Then the operator to be sandwiched between the meson states reads
\begin{multline}
\mel{\vb*{k},\lambda}{\mathcal{H}^{int}_\text{1st}}{0} = -\frac{1}{\sqrt{2 k_0}} \sum_{\alpha=1,2} e_\alpha \sqrt{\frac{M_\alpha + \widehat{E}_\alpha}{2 \widehat{E}_\alpha}} \\
\left[e^{-i\vb*{k}\vdot\widehat{\vb*{r}}_\alpha}\frac{\widehat{\vb*{p}}_\alpha}{M_\alpha+\widehat{E}_\alpha} + \frac{\widehat{\vb*{p}}_\alpha}{M_\alpha+\widehat{E}_\alpha}e^{-i\vb*{k}\vdot\widehat{\vb*{r}}_\alpha} -i \vb*{\sigma}\!_{\alpha} \cp \pqty{e^{-i\vb*{k}\vdot\widehat{\vb*{r}}_\alpha} \frac{\widehat{\vb*{p}}_\alpha}{M_\alpha+\widehat{E}_\alpha} - \frac{\widehat{\vb*{p}}_\alpha}{M_\alpha+\widehat{E}_\alpha} e^{-i\vb*{k}\vdot\widehat{\vb*{r}}_\alpha}}\right] \\
\vdot (  \vb*{\epsilon}_{\vb*{k}}^{\lambda})^{*} \sqrt{\frac{M_\alpha + \widehat{E}_\alpha}{2 \widehat{E}_\alpha}},
\label{comop}
\end{multline}
\end{widetext}
where we have put a hat above $E$ and $\vb*{p}$ and $\vb*{r}$ to make more clear that they are operators, in contrast to $\vb*{k}$ which is a vector number.

It is a common practice in the analysis of radiative transitions in heavy quarkonium to proceed to a nonrelativistic reduction of this operator up to $\frac{\vb*{p}_{c}}{M_{c}}$ order. This has been justified for the use of the nonrelativistic Schr\"{o}dinger equation for the calculation of the states, see for example \cite{LeY88}. However, as the effectiveness of the quark model parameters and the implementation of the experimental masses in the calculation of radiative decays may incorporate in an effective manner relativistic effects the restriction of the operator to the $\frac{\vb*{p}_{c}}{M_{c}}$ order may be questionable, at least for charmonium where the speed of the quarks in the low lying states is not much smaller than the speed of light in vacuum. Therefore we proceed next to the evaluation of the transition amplitude $\mathcal{M}_{J_{F},m_{F},J_{I},m_{I}}^{\lambda}$ , first in the $\frac{\vb*{p}_{c}}{M_{c}}$ approximation and then to all $\frac{\vb*{p}_{c}}{M_{c}}$ orders. As we shall see, the consideration of the higher $\frac{\vb*{p}_{c}}{M_{c}}$ orders becomes essential for an accurate description of the decay widths.

\section{The p/m approximation\label{SIV}}

The $\frac{\vb*{p}_{c}}{M_{c}}$ approximation is defined from \eqref{comop} in the limit $\widehat{E}_{\alpha}=M_{\alpha}$. A thorough analysis of this approximation has been carried out in \cite{Br3-19}. Here we only give the final expressions for the calculation of the amplitude:
\begin{equation}
\mathcal{M}_{J_{F},m_{F},J_{I},m_{I}}^{\lambda}=\sqrt{2M_{I}}\sqrt{2E_{F}}\sum_{\alpha=1,2}\frac{e_{\alpha}}{2M_{c}}\mel{\Psi_{F}}{\mathcal{O}_{\alpha}}{\Psi_{I}},
\label{p/mamp}
\end{equation}
where
\begin{equation}
\ket{\Psi} \equiv \ket{J,m,nL,s}
\label{wf}
\end{equation}
stands for the $c\overline{c}$ spectroscopic $n\,{^{2s+1}\!L_{J}}$ state and $\mathcal{O}_{\alpha}$ for an operator that we detail next.

Thus, for $^{3}\!S_{1}\rightarrow\gamma\,{^{3}\!P_{J}}$ transitions we have (here we write only the electric part of the operator; the form for the magnetic part can be found in the appendices of \cite{Br3-19})
\begin{equation}
\begin{split}
&\langle \mathcal{O}_{\alpha}^{el}\rangle _{\,{^{3}\!S_{1}}\rightarrow\gamma\,{^{3}\!P_{J}}} \equiv \mel{\Psi_{F(  {^{3}\!P_{J}})}}{\mathcal{O}_{\alpha}^{el}}{\Psi_{I(  \,{^{3}\!S_{1}})  }} \\
&= \mel{\Psi_{F(  {^{3}\!P_{J}})}}{e^{i(-1)  ^{\alpha}(  \frac{\vb*{k}\vdot\widehat{\vb*{r}}}{2})}(-1)^{\alpha}2\widehat
{\vb*{p}}\vdot(  \vb*{\epsilon}_{\vb*{k}}^{\lambda})^*}{\Psi_{I(  \,{^{3}\!S_{1}})  }}
\end{split}
\label{Osp0}
\end{equation}
where $\widehat{\vb*{r}}=\frac{\widehat{\vb*{r}}_{1}-\widehat{\vb*{r}}_{2}}{2}$ is the relative position operator and $\widehat{\vb*{p}}=\frac{\widehat{\vb*{p}}_{1}-\widehat{\vb*{p}}_{2}}{2}$ is the relative three-momentum operator.

By using the equality
\begin{equation}
\widehat{\vb*{p}}=-i\frac{M_{c}}{2} \comm{\widehat{\vb*{r}}}{H}
\label{conm}
\end{equation}
where $H$ is the Cornell spectroscopic Hamiltonian, and introducing a Parseval identity in terms of a complete set of intermediate eigenstates $\Bqty{ \ket{\Psi_{int}}}  $ of $H$, the mass dependence in the matrix element can be explicitly extracted:
\begin{multline}
\langle \mathcal{O}_{\alpha}^{el}\rangle _{\,{^{3}\!S_{1}}\rightarrow\gamma\,{^{3}\!P_{J}}}=\\
=-i M_{c}\sum_{int}\mel{\Psi_{F(  {^{3}\!P_{J}})}}{e^{i(-1)  ^{\alpha}(\frac{\vb*{k} \vdot\widehat{\vb*{r}}}{2})}}{\Psi_{int}} (M_{I}-M_{int}) \\
 \mel{\Psi_{int}}{(-1)^{\alpha}\widehat{\vb*{r}}\vdot(  \vb*{\epsilon}_{\vb*{k}}^{\lambda})^{*}}{\Psi_{I(  \,{^{3}\!S_{1}})  }}.
\label{Osp}
\end{multline}
This permits the implementation of the experimental mass differences $(M_{I}-M_{int})$ instead of the calculated ones so that the quark model wavefunctions can be directly tested.

As for $^{3}\!P_{J}\rightarrow\gamma\,{^{3}\!S_{1}}$ transitions, using $\comm{\widehat{\vb*{p}}}{e^{-i\vb*{k}\vdot\widehat{\vb*{r}}}}
=-\vb*{k}e^{-i\vb*{k}\vdot\widehat{\vb*{r}}}$ we get in a similar manner
\begin{multline}
\langle \mathcal{O}_{\alpha}^{el}\rangle _{{^{3}\!P_{J}}\rightarrow\gamma\,{^{3}\!S_{1}}}\equiv \mel{\Psi_{F(  \,{^{3}\!S_{1}})}}{ \mathcal{O}_{\alpha}^{el}}{\Psi_{I({^{3}\!P_{J}})}} \\
= \mel{\Psi_{F(  \,{^{3}\!S_{1}})}}{(-1)  ^{\alpha}2\widehat{\vb*{p}}\vdot(\vb*{\epsilon}_{\vb*{k}}^{\lambda})  ^{*}e^{i(-1)^{\alpha}(\frac{\vb*{k} \vdot\widehat{\vb*{r}}}{2})}}{\Psi_{_{I(  {^{3}\!P_{J}})  }}}\\
=-i M_{c}\sum_{int}(  M_{int}-M_{F})  \mel{\Psi_{F(\,{^{3}\!S_{1}})}}{(-1)^{\alpha}\widehat{\vb*{r}}\vdot(\vb*{\epsilon}_{\vb*{k}}^{\lambda})  ^{*}}{\Psi_{int}}\\
 \mel{\Psi_{int}}{e^{i(-1)^{\alpha}(\frac{\vb*{k} \vdot\widehat{\vb*{r}}}{2})}}{\Psi_{I({^{3}\!P_{J}})}}.
\label{Ops}
\end{multline}

These expressions get further simplified in the so called Long Wavelength (LWL) limit, corresponding to take $e^{i(-1)^{\alpha}(\frac{\vb*{k} \vdot\widehat{\vb*{r}}}{2})}=1$:
\begin{multline}
\langle (  \mathcal{O}_{\alpha}^{el})  _{LWL}\rangle_{\,{^{3}\!S_{1}}\rightarrow\gamma\,{^{3}\!P_{J}}}=\\
-i M_{c}(M_{I}-M_{F}) \mel{\Psi_{F(  {^{3}\!P_{J}})  }}{(\vb*{\epsilon}_{\vb*{k}}^{\lambda})^{*}\vdot(-1)^{\alpha}\widehat{\vb*{r}}}{\Psi_{I(  \,{^{3}\!S_{1}})}}
\label{LWL1},
\end{multline}
\begin{multline}
\langle (  \mathcal{O}_{\alpha}^{el})  _{LWL}\rangle_{{^{3}\!P_{J}}\rightarrow\gamma\,{^{3}\!S_{1}}}=\\
-i M_{c}(  M_{I}-M_{F})\mel{\Psi_{F(  \,{^{3}\!S_{1}})}}{(\vb*{\epsilon}_{\vb*{k}}^{\lambda})^{*}\vdot(-1)^{\alpha}\widehat{\vb*{r}}}{\Psi_{I(  {^{3}\!P_{J}})}}
\label{LWL2}.
\end{multline}
Actually this is the limit commonly used in the literature \cite{Eic08} for the calculation of radiative decays despite the fact that it can only be taken for granted when the condition
\begin{equation}
\abs{\vb*{k}} (  2 \langle \widehat{\vb*{r}}^{2}\rangle^{\frac{1}{2}}) _{F}<1
\label{criterium}
\end{equation}
is satisfied \cite{Br3-19}. Indeed, regarding the considered transitions in charmonium, the only LWL processes are ${2{^{3}\!S_{1}}\rightarrow\gamma^{3}1P_{1,2}}$ as can be checked from Table~\ref{tabLWL} where the root mean square radii have been obtained from Model~II (the same conclusion comes out from Model~I).
\begin{table}
\begin{ruledtabular}
\begin{tabular}{lcc}
$^{3}\!S_{1}\rightarrow\gamma\,{^{3}\!P_{J}}$ & $\abs{\vb*{k}}_\expt\text{(MeV)}$ & $\abs{\vb*{k}}_\expt(  2\left\langle r^{2}\right\rangle ^{\frac{1}{2}})_{{^{3}\!P_{J}}}$\\
\hline
$\psi(2S)  \rightarrow\gamma\chi_{c_{0}}(  1p)  $ &
$261$ & $1.6$\\
$\psi(2S)  \rightarrow\gamma\chi_{c_{1}}(  1p)  $ &
$171$ & $1.0$\\
$\psi(2S)  \rightarrow\gamma\chi_{c_{2}}(  1p)  $ &
$128$ & $0.8$\\
$\chi_{c_{0}}(1p)  \rightarrow\gamma J/\psi$ & $303$ & $1.2$\\
$\chi_{c_{1}}(1p)  \rightarrow\gamma J/\psi$ & $389$ & $1.5$\\
$\chi_{c_{2}}(1p)  \rightarrow\gamma J/\psi$ & $430$ & $1.6$\\
\end{tabular}
\end{ruledtabular}
\caption{Experimental values of the photon energy $\abs{\vb*{k}}_\expt$ and calculated values of $\abs{\vb*{k}}_\expt(  2\left\langle r^{2}\right\rangle ^{\frac{1}{2}})  _{{^{3}\!P_{J}}}$ from Model~II for $^{3}\!S_{1}\rightarrow\gamma\,{^{3}\!P_{J}}$ and $^{3}\!P_{J}\rightarrow\gamma\,{^{3}\!S_{1}}$ radiative transitions.}
\label{tabLWL}
\end{table}

In Table~\ref{tabp/m} we list the calculated ${{^{3}\!S_1} \leftrightarrow {^{3}\!P_J}}$ transition widths as compared to data for charmonium. In all cases the experimental masses for the initial, final and (when known) intermediate charmonium states instead of the calculated ones from the spectroscopic Hamiltonian have been used. It turns out that the consideration of intermediate $n{^{3}\!P_{J}}$ states with $n\leq2$ is enough in the sense that the inclusion of higher Cornell states hardly changes ($2\%$ at most) the results. Regarding the intermediate $2{^{3}\!P_{J}}$ states we have used the experimental masses for $2{^{3}\!P_{0}}$, corresponding to $\chi_{c0}(3860)$ under the assumption that this resonance is a pure Cornell state, and $2{^{3}\!P_{2}}$, corresponding to $\chi_{c2}(  3930)$ that may be reasonably taken as a pure Cornell state since it lies quite below the first $S$-wave $2^{++}$ meson-meson threshold \cite{Gon15}. As for $2{^{3}\!P_{1}}$ we have used the calculated mass from Model~I, see Table~\ref{tabIII}, since it lies in between those of $2{^{3}\!P_{0}}$ and $2{^{3}\!P_{2}}$ as should be expected when no threshold effects are taken into account (notice that the $X(3872)$ can not be taken as a pure Cornell state).

\begin{table*}
\begin{ruledtabular}
\begin{tabular}{lccccc}
Radiative Decay & $(\Gamma_{LWL}^{(  \theor-\expt)})_{I}$ & $(\Gamma_{p/M}^{(  \theor-\expt)})_{I}$ & $\Gamma_\expt^{PDG}$ & $(\Gamma_{p/M}^{(\theor-\expt)})_{II}$ & $(\Gamma_{LWL}^{(\theor-\expt)})_{II} $ \\
& KeV & KeV & KeV & KeV & KeV \\
\hline
$\psi(  2S)  \rightarrow\gamma\chi_{c_{0}}(  1p)  $ &
$61$ & $77$ & $28.8\pm1.4$ & $57$ & $47$\\
$\psi(  2S)  \rightarrow\gamma\chi_{c_{1}}(  1p)  $ &
$53$ & $52$ & $28.7\pm1.5$ & $41$ & $41$\\
$\psi(  2S)  \rightarrow\gamma\chi_{c_{2}}(  1p)  $ &
$37$ & $37$ & $28.0\pm1.3$ & $29$ & $29$\\
$\chi_{c_{0}}(  1p)  \rightarrow\gamma J/\psi$ & $186$ & $160$ &
$151\pm14$ & $118$ & $128$\\
$\chi_{c_{1}}(  1p)  \rightarrow\gamma J/\psi$ & $386$ & $464$ &
$288\pm22$ & $315$ & $266$\\
$\chi_{c_{2}}(  1p)  \rightarrow\gamma J/\psi$ & $513$ & $616$ &
$374\pm27$ & $419$ & $353$ \\
\end{tabular}
\end{ruledtabular}
\caption{Calculated widths up to order $\frac{\vb*{p}_{c}}{M_{c}}$ as compared to data for $\psi(  2S)  \rightarrow\gamma\chi_{c_{J}}(  1P)$ and $\chi_{c_{J}}(1P)\rightarrow\gamma J/\psi$. Notation as follows. $\Gamma_{LWL}^{(\theor-\expt)}$: width in the LWL approximation implemented with the experimental masses and photon energy. $\Gamma_{p/M}^{(\theor-\expt)}$: width in the $\frac{\vb*{p}_{c}}{M_{c}}$ approximation implemented with the experimental masses and photon energy. The subscripts $I$ and $II$ refer to Model~I and II. $\Gamma_\expt^{PDG}$: measured widths \cite{PDG18}.}
\label{tabp/m}
\end{table*}

A look at Table~\ref{tabp/m} confirms, through a comparison of the calculated LWL results (second column for Model~I and sixth column for Model~II) with the $\frac{\vb*{p}_{c}}{M_{c}}$ ones (third column for Model~I and fifth column for Model~II), the validity of the LWL limit for $2{^{3}\!S_{1}}\rightarrow\gamma^{3}1P_{1,2}$.

It makes also clear, through the comparison of the calculated $\frac{\vb*{p}_{c}}{M_{c}}$ decay widths with data, that the $\frac{\vb*{p}_{c}}{M_{c}}$ approximation does not give an accurate overall description of these decays. More precisely, except for $1{^{3}\!P_{0}}\rightarrow\gamma\,1{^{3}\!S_{1}}$ from Model~I and $2{^{3}\!S_{1}}\rightarrow\gamma 1{^{3}\!P_{2}}$ from Model~II, all the calculated $\frac{\vb*{p}_{c}}{M_{c}}$ widths are out of the experimental intervals, in some cases with big differences respect to data. This is in contrast with the situation in bottomonium \cite{Br3-19} where most of the calculated widths were within or pretty close to the measured intervals.

By realizing that this may have to do at least in part with the poor convergence of the expansion of the complete transition operator in powers of $\frac{\vb*{p}_{c}}{M_{c}}$ (let us recall that the expectation value of $\frac{\left\vert \vb*{p}_{c}\right\vert }{M_{c}}$ in the low lying charmonium states can be as big as $0.5$) we develop in what follows the formalism for the application of the complete operator \eqref{comop} to the calculation of the decay widths.

\section{Beyond the p/m approximation\label{SV}}

As it was mentioned before, the extraction of the mass dependence in the matrix elements involved in the calculation of the radiative decay widths, allowing for the substitution of the calculated masses by the measured ones, is a crucial step to test the spectroscopic quark model wavefunctions. Moreover, it is a condition \emph{sine qua non} to get an accurate description of radiative decay widths in bottomonium \cite{Br3-19}.

When the complete transition operator \eqref{comop} is considered, the presence of the momentum dependent operator $\widehat{E}_{\alpha}$ (instead of the constant $M_{\alpha}$ in the $\frac{\vb*{p}_{c}}{M_{c}}$ approximation) complicates this mass extraction. To facilitate it, we first rearrange expression \eqref{comop} to group all the energy dependent operators to the right so that they act on the initial state. This is convenient on the one hand because we can extract the mass dependence in the terms multiplying the energy dependent operators in exactly the same manner as done in the $\frac{\vb*{p}_{c}}{M_{c}}$ approximation (see below), and on the other hand because the action of $\widehat{E}_{\alpha}$ simplifies when acting on a state of zero total three-momentum. In fact, as the initial state is in its rest frame the action of $\widehat{\vb*{p}}_{\alpha}=\frac{\widehat{\vb*{P}}}{2}-(-1)^{\alpha}\widehat{\vb*{p}}$ becomes equivalent to that of $\widehat{\vb*{p}}$:
\begin{multline}
\widehat{E}_{\alpha}\ket{\vb*{P=0,}J,m,nL,s} =\\
\sqrt{M_{c}^{2}+\widehat{\vb*{p}}^{2}}\ket{\vb*{P=0},J,m,nL,s}.
\label{P=0}
\end{multline}

To relocate the energy dependent operators we use that $e^{-i\vb*{k}\vdot\widehat{\vb*{r}}_{\alpha}}$ represents a three-momentum translation, so that
\begin{equation}
e^{-i\vb*{k}\vdot\widehat{\vb*{r}}_{\alpha}}\ket{\vb*{p}_{\alpha}} =\ket{\vb*{p}_{\alpha}-\vb*{k}}.
\label{trans1}
\end{equation}
If we expand for convenience the meson states in terms of the quark and antiquark momentum eigenstates, $\ket{\vb*{p}_1,\vb*{p}_2}$ where we obviate the spin quantum number since it does not play any role in this argument, then \eqref{trans1} gives account of the transition from a quark (antiquark) state with three-momentum $\vb*{p}_{\alpha}$ to a quark (antiquark) state with three-momentum $(\vb*{p}_{\alpha}-\vb*{k})$ through the emission of a photon with three-momentum $\vb*{k}$. Then energy conservation tells us that
\begin{equation}
\widehat{E}_{\alpha}(e^{-i\vb*{k}\vdot\widehat{\vb*{r}}_{\alpha}} \ket{\vb*{p}_{\alpha}})  = \widehat{E}_{\alpha} \ket{\vb*{p}_{\alpha}-\vb*{k}} =(E_{\alpha}-k_{0}) \ket{\vb*{p}_{\alpha}-\vb*{k}},
\label{trans2}
\end{equation}
where $E_{\alpha}$ and $k_{0}$ are numbers, not operators, so that we can write
\begin{equation}
\begin{split}
\widehat{E}_{\alpha}(e^{-i\vb*{k}\vdot\widehat{\vb*{r}}_{\alpha}}\ket{\vb*{p}_{\alpha}}) &=  (  E_{\alpha}-k_{0}) \pqty*{e^{-i\vb*{k}\vdot\widehat{\vb*{r}}_{\alpha}} \ket{\vb*{p}_{\alpha}}} \\
&= e^{-i\vb*{k}\vdot\widehat{\vb*{r}}_{\alpha}}(  E_{\alpha}-k_{0})  \ket{\vb*{p}_{\alpha}} \\
&=e^{-i\vb*{k}\vdot\widehat{\vb*{r}}_{\alpha}}(  \widehat{E}_{\alpha}-k_{0}) \ket{ \vb*{p}_{\alpha}}.
\end{split}
\label{trans3}
\end{equation}
Since this equality holds for the complete set of momentum eigenstates  $\Bqty{\ket{\vb*{p}_\alpha}}$ we can rewrite it as an identity between operators:
\begin{equation}
\widehat{E}_{\alpha}e^{-i\vb*{k}\vdot\widehat{\vb*{r}}_{\alpha}}=e^{-i\vb*{k}\vdot\widehat{\vb*{r}}_{\alpha}}(  \widehat{E}_{\alpha}-k_{0}).
\label{trans4}
\end{equation}
Then, using \eqref{trans4} and the well known commutator
\begin{equation}
\comm{\widehat{\vb*{p}}_{\alpha}}{e^{-i\vb*{k}\vdot\widehat{\vb*{r}}_{\alpha}}}  =-\vb*{k}e^{-i\vb*{k}\vdot\widehat{\vb*{r}}_{\alpha}}
\label{conmpk}
\end{equation}
we rearrange \eqref{comop} as
\begin{widetext}
\begin{multline}
\mel{\vb*{k},\lambda }{ \mathcal{H}_\text{1st}^{int}}{0}  =-\frac{1}{\sqrt{2k_{0}}}\sum_{\alpha=1,2}\frac{e_{\alpha}}{2M_{\alpha}}\{e^{-i\vb*{k}\vdot\widehat{\vb*{r}}_{\alpha}}2\widehat{\vb*{p}}_{\alpha}\;\mathcal{P}_{+}(\widehat{E}_{\alpha})-e^{-i\vb*{k}\vdot\widehat{\vb*{r}}_{\alpha}}\vb*{k}\;\mathcal{K}(\widehat{E}_{\alpha})+\\
 -i\vb*{\sigma}_{\alpha}\times\lbrack e^{-i\vb*{k}\vdot\widehat{\vb*{r}}_{\alpha}}2\widehat{\vb*{p}}_{\alpha}\;\mathcal{P}_{-}(\widehat{E}_{\alpha})+e^{-i\vb*{k}\vdot\widehat{\vb*{r}}_{\alpha}}\vb*{k}\;\mathcal{K}(\widehat{E}_{\alpha})]\}\vdot(\vb*{\epsilon}_{\vb*{k}}^{\lambda})  ^{*}
\label{2.4}
\end{multline}
where $\mathcal{P}_{\pm}(\widehat{E}_{\alpha})$ and $\mathcal{K}(\widehat{E}_{\alpha})$ stand for the energy dependent scalar operators
\begin{equation}
\mathcal{P}_{\pm}(\widehat{E}_{\alpha}) \equiv \pqty{\frac{M_{\alpha}}{M_{\alpha}+\widehat{E}_{\alpha}}\pm\frac{M_{\alpha}}{M_{\alpha}+\widehat{E}_{\alpha}-k_{0}}} \sqrt{\frac{(  M_{\alpha}+\widehat{E}_{\alpha})(  M_{\alpha}+\widehat{E}_{\alpha}-k_{0})}{4\widehat{E}_{\alpha}(  \widehat{E}_{\alpha}-k_{0})}},
\label{pen}
\end{equation}
\begin{equation}
\mathcal{K}(\widehat{E}_{\alpha}) \equiv\pqty{\frac{2M_{\alpha}}{M_{\alpha}+\widehat{E}_{\alpha}-k_{0}}} \sqrt{\frac{(  M_{\alpha}+\widehat{E}_{\alpha})(M_{\alpha}+\widehat{E}_{\alpha}-k_{0})  }{4\widehat{E}_{\alpha}(  \widehat{E}_{\alpha}-k_{0})  }}.
\label{ken}
\end{equation}

Equivalently, using \eqref{conmpk} we can also write
\begin{multline}
\mel{\vb*{k},\lambda}{\mathcal{H}^{int}_\text{1st}}{0}  =-\frac{1}{\sqrt{2k_{0}}}\sum_{\alpha=1,2}\frac{e_{\alpha}}{2M_{\alpha}}\{2\widehat{\vb*{p}}_{\alpha}e^{-i\vb*{k}\vdot\widehat{\vb*{r}}_{\alpha}}\mathcal{P}_{+}(\widehat{E}_{\alpha})+e^{-i\vb*{k}\vdot\widehat{\vb*{r}}_{\alpha}}\vb*{k}\;\mathcal{K}'(\widehat{E}_{\alpha})+\\
  -i\vb*{\sigma}_{\alpha}\times\lbrack2\widehat{\vb*{p}}_{\alpha}e^{-i\vb*{k}\vdot\widehat{\vb*{r}}_{\alpha}}\mathcal{P}_{-}(\widehat{E}_{\alpha})+e^{-i\vb*{k}\vdot\widehat{\vb*{r}}_{\alpha}}\vb*{k}\;\mathcal{K}'(\widehat{E}_{\alpha})]\}\vdot(  \vb*{\epsilon}_{\vb*{k}}^{\lambda})^{*},
\label{2.5}
\end{multline}
\end{widetext}
with 
\begin{equation}
\mathcal{K}'(\widehat{E}_{\alpha}) \equiv\pqty{\frac{2M_{\alpha}}{M_{\alpha}+\widehat{E}_{\alpha}}} \sqrt{\frac{(  M_{\alpha}+\widehat{E}_{\alpha})(M_{\alpha}+\widehat{E}_{\alpha}-k_{0})  }{4\widehat{E}_{\alpha}(  \widehat{E}_{\alpha}-k_{0})  }}.
\label{kprimen}
\end{equation}
For ${{^{3}\!S_1} \leftrightarrow {^{3}\!P_J}}$ transitions, the use of \eqref{2.4} makes easier the calculation when the initial state is an $S$-wave ${(L_{I}=0)}$, whereas \eqref{2.5} is more convenient when the initial state is a $P$-wave $(L_{I}=1)$, as it was the case in the  $\frac{\vb*{p}}{M}$ approximation \cite{Br3-19}.

From \eqref{2.4} or \eqref{2.5} the transition amplitude $\mathcal{M}_{J_{F},m_{F},J_{I},m_{I}}^{\lambda}$ can be straightforwardly derived following the step by step procedure explained in \cite{Br3-19}. Thus, using
\begin{align}
\ket{I} &= \ket{\vb*{P}_{I},J_{I},m_{I},n_{I}L_{I},s_{I}},
\label{IN} \\
\ket{F} &= \ket{\vb*{P}_{F},J_{F},m_{F},n_{F}L_{F},s_{F}},
\label{FIN}
\end{align}
introducing center of mass
\begin{equation}
\widehat{\vb*{R}}=\frac{\widehat{\vb*{r}}_{1}+\widehat{\vb*{r}}_{2}}{2} \qquad \widehat{\vb*{P}}=\widehat{\vb*{p}}_{1}+\widehat{\vb*{p}}_{2}
\label{cm}
\end{equation}
and relative
\begin{equation}
\widehat{\vb*{r}}=\widehat{\vb*{r}}_{1}-\widehat{\vb*{r}}_{2} \qquad \widehat{\vb*{p}}=\frac{\widehat{\vb*{p}}_{1}-\widehat
{\vb*{p}}_{2}}{2}
\label{rel}
\end{equation}
operators, integrating over the center of mass spatial degrees of freedom, taking into account that ${(\vb*{\epsilon}_{\vb*{k}}^{\lambda}) ^{*}\vdot\vb*{k}=0}$ and that in the rest frame of the decaying meson one has $\vb*{P}_{I}=\vb*{0}$, $\vb*{P}_{F}=-\vb*{k}$, the transition amplitude becomes
\begin{equation}
\mathcal{M}_{J_{F},m_{F},J_{I},m_{I}}^{\lambda}=\sqrt{2M_{I}}\sqrt{2E_{F}}\sum_{\alpha=1,2}\frac{e_{\alpha}}{2M_{\alpha}}\mel{\Psi_{F}}{\widetilde{\mathcal{O}}_{\alpha}}{\Psi_{I}}
\label{tmec}
\end{equation}
where the matrix element $\mel{\Psi_{F}}{\widetilde{\mathcal{O}}_{\alpha}}{\Psi_{I}} \equiv \langle \widetilde{\mathcal{O}}_{\alpha}\rangle _{FI}$ can be conveniently expressed as a sum of electric and magnetic contributions
\begin{equation}
\begin{split}
\langle \widetilde{\mathcal{O}}_{\alpha}\rangle_{FI}  &=\langle \widetilde{\mathcal{O}}_{\alpha}\rangle_{FI}^{el}+\langle\widetilde{\mathcal{O}}_{\alpha}\rangle _{FI}^{(  mag)_{\vb*{\sigma}\cp\vb*{k}}}+\langle \widetilde{\mathcal{O}}_{\alpha}\rangle_{FI}^{(  mag)  _{\vb*{\sigma}\cp\widehat{\vb*{p}}}}\\
&=\langle \widetilde{\mathcal{O}}_{\alpha}^{\prime}\rangle_{FI}^{el}+\langle\widetilde{\mathcal{O}}_{\alpha}^{\prime}\rangle _{FI}^{(mag)  _{\vb*{\sigma}\cp\vb*{k}}}+\langle \widetilde{\mathcal{O}}_{\alpha}^{\prime}\rangle_{FI}^{(  mag)  _{\vb*{\sigma}\cp\widehat{\vb*{p}}}},
\end{split}
\label{reesp}
\end{equation}
the first decomposition is technically convenient when the initial state is an $S$-wave, the second one when the initial state is a $P$-wave.

More explicitly, in the first decomposition the electric part is given by
\begin{multline}
\langle \widetilde{\mathcal{O}}_{\alpha}\rangle_{FI}^{el}=\\
\mel{\Psi_{F}}{e^{i(-1)^{\alpha}(\frac{\vb*{k}\vdot\widehat{\vb*{r}}}{2})}(-1)^{\alpha}2\widehat{\vb*{p}}\vdot(  \vb*{\epsilon}_{\vb*{k}}^{\lambda})^{*}\mathcal{P}_{+}(\widehat{E}_{\alpha})}{\Psi_{I}},
\label{Oel}
\end{multline}
the first magnetic term by
\begin{multline}
\langle \widetilde{\mathcal{O}}_{\alpha}\rangle_{FI}^{(mag)_{\vb*{\sigma}\cp\vb*{k}}}=\\
\mel{\Psi_{F}}{e^{i(-1)  ^{\alpha}(  \frac{\vb*{k}\vdot\widehat{\vb*{r}}}{2})  }i\vb*{\sigma}_{\alpha}\times\vb*{k}\vdot(\vb*{\epsilon}_{\vb*{k}}^{\lambda})^{*}\mathcal{K}(\widehat{E}_{\alpha})}{\Psi_{I}}
\label{Omag}
\end{multline}
and the second magnetic term by
\begin{multline}
\langle \widetilde{\mathcal{O}}_{\alpha}\rangle_{FI}^{(mag)_{\vb*{\sigma}\cp\widehat{\vb*{p}}}}=\\
-\mel{\Psi_{F}}{e^{i(-1)^{\alpha}(\frac{\vb*{k}\vdot\widehat{\vb*{r}}}{2})}i\vb*{\sigma}_{\alpha}\times(-1)^{\alpha}2\widehat{\vb*{p}}\vdot(  \vb*{\epsilon}_{\vb*{k}}^{\lambda})^{*}\mathcal{P}_{-}(\widehat{E}_{\alpha})}{\Psi_{I}}.
\label{Omagp}
\end{multline}
As can be easily checked these expressions reduce to the corresponding ones obtained in the $\frac{\vb*{p}_{c}}{M_{c}}$ approximation by taking $\widehat{E}_{\alpha}=M_{\alpha}$ and $k_{0}\ll M_{\alpha}$.

\bigskip

The second decomposition reads
\begin{multline}
\langle \widetilde{\mathcal{O}}_{\alpha}^{\prime}\rangle_{FI}^{el} = \\
\mel{\Psi_{F}}{(-1)^{\alpha}2\widehat{\vb*{p}} \vdot(\vb*{\epsilon}_{\vb*{k}}^{\lambda})^{*}e^{i(-1)^{\alpha}(\frac{\vb*{k}\vdot\widehat{\vb*{r}}}{2})  }\mathcal{P}_{+}(\widehat{E}_{\alpha})}{\Psi_{I}},
\label{OpLi1}
\end{multline}
\begin{multline}
\langle \widetilde{\mathcal{O}}_{\alpha}^{\prime}\rangle_{FI}^{(  mag)_{\vb*{\sigma}\cp\vb*{k}}} = \\
\mel{\Psi_{F}}{i\vb*{\sigma}_{\alpha}\times\vb*{k}\vdot(\vb*{\epsilon}_{\vb*{k}}^{\lambda})^{*}e^{i(-1)^{\alpha}(\frac{\vb*{k}\vdot\widehat{\vb*{r}}}{2})}\mathcal{P}_{+}(\widehat{E}_{\alpha})}{\Psi_{I}},
\label{OpLi1mag}
\end{multline}
\begin{multline}
\langle \widetilde{\mathcal{O}}_{\alpha}^{\prime}\rangle_{FI}^{(  mag)_{\vb*{\sigma}\cp\widehat{\vb*{p}}}} = \\
-\mel{\Psi_{F}}{i\vb*{\sigma}_{\alpha}\times(-1)^{\alpha}2\widehat{\vb*{p}}\vdot(\vb*{\epsilon}_{\vb*{k}}^{\lambda})^{*}e^{i(-1)^{\alpha}(\frac{\vb*{k} \vdot\widehat{\vb*{r}}}{2})}\mathcal{P}_{-}(\widehat{E}_{\alpha})}{\Psi_{I}}.
\label{Omagprimep}
\end{multline}

In order to extract the mass dependence from the $\widehat{\vb*{p}}$ operator we use \eqref{conm} and introduce two Parseval identities in terms of eigenstates of the Cornell Hamiltonian
\begin{equation}
I = \sum_{A}\ketbra{\Psi_{A}}{\Psi_{A}}
\label{Parsid}
\end{equation}
where $A$ is a shorthand notation for all the quantum numbers labeling the eigenstates ($J_A,m_A,n_A, L_A,s_A$), so that the above expressions become
\begin{widetext}
\begin{gather}
\langle \widetilde{\mathcal{O}}_{\alpha}\rangle_{FI}^{el}  =
-iM_{c}\sum_{A,B}\mel{\Psi_{F}}{e^{i(-1)^{\alpha}(\frac{\vb*{k}\vdot\widehat{\vb*{r}}}{2})}}{\Psi_{A}}(M_{B}-M_{A}) \mel{\Psi_{A}}{(-1)  ^{\alpha}\widehat{\vb*{r}}\vdot(\vb*{\epsilon}_{\vb*{k}}^{\lambda})  ^{*}}{\Psi_{B}} \mel{\Psi_{B}}{\mathcal{P}_{+}(\widehat{E}_{\alpha})}{\Psi_{I}}, \label{OelFI}\\
\langle \widetilde{\mathcal{O}}_{\alpha}\rangle_{FI}^{(mag)_{\vb*{\sigma}\cp\vb*{k}}}=
\sum_{B}\mel{\Psi_{F}}{e^{i(-1)  ^{\alpha}(\frac{\vb*{k} \vdot\widehat{\vb*{r}}}{2})  }i\vb*{\sigma}_{\alpha}\times \vb*{k}  \vdot(\vb*{\epsilon}_{\vb*{k}}^{\lambda})  ^{*}}{\Psi_{B}}
\mel{\Psi_{B}}{\mathcal{K}(\widehat{E}_{\alpha})}{\Psi_{I}}, \label{OmagkFI}\\
\langle \widetilde{\mathcal{O}}_{\alpha}\rangle_{FI}^{(mag)_{\vb*{\sigma}\cp\widehat{\vb*{p}}}}=
iM_{c}\sum_{A,B}\mel{\Psi_{F}}{e^{i(-1)  ^{\alpha}(\frac{\vb*{k}\vdot\widehat{\vb*{r}}}{2})  }}{\Psi_{A}} (M_{B}-M_{A}) 
\mel{\Psi_{A}}{(-1)^{\alpha}i\vb*{\sigma}_{\alpha}\times\widehat{\vb*{r}}\vdot(  \vb*{\epsilon}_{\vb*{k}}^{\lambda})^{*}}{\Psi_{B}} \mel{\Psi_{B}}{\mathcal{P}_{-}(\widehat{E}_{\alpha})}{\Psi_{I}}, \label{OmagpFI} 
\end{gather}
and
\begin{gather}
\langle \widetilde{\mathcal{O}}_{\alpha}^{\prime}\rangle_{FI}^{el}  = -iM_{c}\sum_{A,B}\mel{\Psi_{F}}{(-1)  ^{\alpha}\widehat{\vb*{r}}\vdot(\vb*{\epsilon}_{\vb*{k}}^{\lambda})  ^{*}}{\Psi_{A}}(M_{A}-M_{F}) \mel{\Psi_{A}}{e^{i(-1)^{\alpha}(\frac{\vb*{k}\vdot\widehat{\vb*{r}}}{2})}}{\Psi_{B}} \mel{\Psi_{B}}{\mathcal{P}_{+}(\widehat{E}_{\alpha})}{\Psi_{I}}, \label{OprimelFI}\\
\langle \widetilde{\mathcal{O}}_{\alpha}^{\prime}\rangle_{FI}^{(mag)_{\vb*{\sigma}\cp\vb*{k}}}= \sum_{B}\mel{\Psi_{F}}{e^{i(-1)  ^{\alpha}(\frac{\vb*{k} \vdot\widehat{\vb*{r}}}{2})  }i\vb*{\sigma}_{\alpha}\times \vb*{k}  \vdot(\vb*{\epsilon}_{\vb*{k}}^{\lambda})  ^{*}}{\Psi_{B}} \mel{\Psi_{B}}{\mathcal{P}_{+}(\widehat{E}_{\alpha})}{\Psi_{I}} , \label{OprimemagkFI} \\
\langle \widetilde{\mathcal{O}}_{\alpha}^{\prime}\rangle_{FI}^{(mag)_{\vb*{\sigma}\cp\widehat{\vb*{p}}}}=
iM_{c}\sum_{A,B}\mel{\Psi_{F}}{(-1)^{\alpha}i\vb*{\sigma}_{\alpha}\times\widehat{\vb*{r}}\vdot(  \vb*{\epsilon}_{\vb*{k}}^{\lambda})^{*}}{\Psi_{A}} (M_{A}-M_{F})  
\mel{\Psi_{A}}{e^{i(-1)  ^{\alpha}(\frac{\vb*{k}\vdot\widehat{\vb*{r}}}{2})  }}{\Psi_{B}} \mel{\Psi_{B}}{\mathcal{P}_{-}(\widehat{E}_{\alpha})}{\Psi_{I}} . \label{OprimemagpFI}
\end{gather}
\end{widetext}

\bigskip

To proceed further we should extract the state mass dependence, if present, from the matrix elements involving the energy dependent operators. Although in principle one could extract it through the expansion of the energy operators in powers of $\frac{\widehat{\vb*{p}}^{2}}{M_{c}^{2}}$ and the introduction of as many additional complete sets of intermediate states as needed, this is not practicable. Instead we can infer the possible mass dependence by realizing that
\begin{enumerate}[i)]
\item $k_{0}$ is quite smaller than $M_{c}$, $\frac{k_{0}}{M_{c}}\leq0.3$ for Model~I and $\frac{k_{0}}{M_{c}}\leq0.23$ for Model~II, as can be verified from Table~\ref{tabLWL}. Then, up to $(  \frac{k_{0}}{M_{c}})  ^{0}$ order the energy dependent operators reduce to (let us recall that as $\widehat{E}_{\alpha}$ is acting on the initial state it can be substituted by $\widehat{E}=\sqrt{M_{c}^{2}+\widehat{\vb*{p}}^{2}}$)
\begin{equation}
\mathcal{P}_{+}(\widehat{E})\simeq\frac{M_{c}}{\widehat{E}}\simeq\mathcal{K}(\widehat{E})
\label{ME}
\end{equation}
\begin{equation}
\mathcal{P}_{-}(\widehat{E})\simeq {0} \label{P-}
\end{equation}
where
\begin{equation}
\frac{M_{c}}{\widehat{E}}=\frac{1}{\sqrt{1+\frac{\widehat{\vb*{p}}^{2}}{M_{c}^{2}}}}=1-\frac{\widehat{\vb*{p}}^{2}}{2M_{c}^{2}}+\frac{3}{2} \pqty{\frac{\widehat{\vb*{p}}^{2}}{2M_{c}^{2}}}^{2}+\dots
\label{M/E}
\end{equation}
and
\begin{equation}
\frac{\widehat{\vb*{p}}^{2}}{2M_{c}^{2}}=\frac{H-V(  r)}{2M_{c}} .
\label{HV}
\end{equation}

\item as the energy dependent operators are scalars, the only contributions to the matrix elements, for example $\mel{\Psi_{B}}{ \mathcal{P}_{+}(\widehat{E}_{\alpha})}{\Psi_{I}}$, come from $\ket{\Psi_{B}}$ being equal to $\ket{\Psi_{I}}$ or to a radial excitation of $\ket{\Psi_{I}}$. The dominant contribution is the diagonal one $\mel{\Psi_{I}}{\mathcal{P}_{+}(\widehat{E}_{\alpha})}{\Psi_{I}}$ involving the matrix element $\mel{\Psi_{I}}{H-V(  r)}{\Psi_{I}} =(M_{I}-2M_{c})  -\mel{\Psi_{I}}{V(  r)}{\Psi_{I}}$.

\item according to our assumption that the difference between the calculated mass and the experimental one is due to potential corrections evaluated at first order in perturbation theory it is obvious that $(M_{I})_\text{calculated}-\mel{\Psi_{I}}{V(  r)}{\Psi_{I}} =(  M_{I})  _\expt-\mel{\Psi_{I}}{V_\text{corrected}(  r)}{\Psi_{I}}$ for any of our quark models. Hence, we can appropriately calculate $\mel{\Psi_{I}}{H-V(r)}{\Psi_{I}}$ with Model~I or II without the need of extracting the mass dependence to implement the measured values. Notice though that the more significant the contribution of the non diagonal matrix element of the energy dependent operators, whose mass dependence has not been extracted, the more uncertain the result.
\end{enumerate}

Therefore, as the consideration of higher $\frac{k_{0}}{M_{c}}$ orders do not introduce any additional dependence on the mass of the states, we can calculate the matrix elements of the energy dependent operators from Model~I or II. One should keep in mind though that the presence of the energy dependent operators, depending on $\frac{\widehat{\vb*{p}}^{2}}{2M_{c}^{2}}=\frac{H-V(  r)}{2M_{c}}$, introduces an additional quark mass dependence in the amplitude with respect to the $\frac{\vb*{p}_{c}}{M_{c}}$ approximation. This makes the calculated widths to be more model dependent so that their comparison to data is not only testing the quark model wavefunctions but also the quark mass parameter. This means that this parameter looses part of its effectiveness becoming a more phenomenological one. In fact, we show next that the analysis of radiative transitions may severely constraint its range of values in correlation with the phenomenological string tension.

For the sake of completeness we write also the former matrix elements in the LWL limit:
\begin{widetext}
\begin{gather}
(  \langle \widetilde{\mathcal{O}}_{\alpha}\rangle_{FI}^{el})  _{LWL}= (\langle \widetilde{\mathcal{O}}_{\alpha}^{\prime}\rangle_{FI}^{el})_{LWL}  =-iM_{c}\sum_{B}\mel{\Psi_{F}}{(  M_{B}-M_{F})  (-1)  ^{\alpha}\widehat{\vb*{r}}\vdot(  \vb*{\epsilon}_{\vb*{k}}^{\lambda})  ^{*}}{\Psi_{B}} \mel{\Psi_{B}}{\mathcal{P}_{+}(\widehat{E}_{\alpha})}{\Psi_{I}};  \label{finlwlel} \\
(  \langle \widetilde{\mathcal{O}}_{\alpha}\rangle_{FI}^{(  mag)  _{\vb*{\sigma}\cp\vb*{k}}})  _{LWL}=(\langle \widetilde{\mathcal{O}}_{\alpha}^{\prime}\rangle_{FI}^{(mag)_{\vb*{\sigma}\cp\vb*{k}}})_{LWL}= \sum_{B}\mel{\Psi_{F}}{ (  i\vb*{\sigma}_{\alpha}\times\vb*{k})\vdot(  \vb*{\epsilon}_{\vb*{k}}^{\lambda})  ^{*}}{\Psi_{B}} \mel{\Psi_{B}}{ \mathcal{K}(\widehat{E}_{\alpha})}{ \Psi_{I}}. \label{finlwlmagk} \\
(\langle \widetilde{\mathcal{O}}_{\alpha}\rangle_{FI}^{(  mag)  _{\vb*{\sigma}\cp\widehat{\vb*{p}}}})  _{LWL}= (\langle \widetilde{\mathcal{O}}_{\alpha}^{\prime}\rangle_{FI}^{(mag)_{\vb*{\sigma}\cp\widehat{\vb*{p}}}})_{LWL} = iM_{c}\sum_{B}\mel{\Psi_{F}}{i\vb*{\sigma}_{\alpha}\times(  M_{B}-M_{F})  (-1)  ^{\alpha}\widehat{\vb*{r}}\vdot(  \vb*{\epsilon}_{\vb*{k}}^{\lambda})  ^{*}}{ \Psi_{B}} \mel{\Psi_{B}}{\mathcal{P}_{-}(\widehat{E}_{\alpha})}{ \Psi_{I}}; \label{finlwlmagp}
\end{gather}

\subsection{Transitions $^{3}\!S_{1}\rightarrow\gamma\,{^{3}\!P_{J}}$}

In order to evaluate the $^{3}\!S_{1}\rightarrow\gamma\,{^{3}\!P_{J}}$ amplitude we use $\widehat{\vb*{r}}\vdot(\vb*{\epsilon}_{\vb*{k}}^{\lambda})  ^{*}=\sqrt{\frac{4\pi}{3}}r(  Y_{1}^{\lambda}(\hat{r})  )  ^{*}$, the well known expansion
\begin{equation}
e^{i(  \frac{\vb*{k}\vdot\vb*{r}}{2})  }=\sum_{l=0}^{\infty} i^{l}\sqrt{4\pi}\sqrt{2l+1}j_{l}\pqty{\frac{kr}{2}}  Y_{l}^{0}(\hat{r})  \label{eikr}
\end{equation}
where $k$ is in the $Z$ direction, and the fact that the energy dependent operators are scalars, so that for example
\begin{equation}
\mel{\Psi_{B}}{\mathcal{P}_{+}(\widehat{E}_{\alpha})}{\Psi_{I}} \propto\delta_{J_{B},J_{I}}\delta
_{m_{B},m_{I}}\delta_{L_{B},L_{I}}\delta_{s_{B},s_{I}} .
\label{deltaBI}
\end{equation}
Thus, from \eqref{tmec} and \eqref{OelFI} the electric part of the amplitude reads
\begin{multline}
(  \mathcal{M}_{J_{F},m_{F},J_{I},m_{I}}^{\lambda\;\text{\text{\text{\text{(electric)}}}}  })  ^{\,{^{3}\!S_{1}}\rightarrow\gamma\,{^{3}\!P_{J}}}=i\sqrt{2M_{I}}\sqrt{2E_{F}}\delta_{S_{I},S_{F}}e_{c} \sum_{l=0}^{\infty}\sum_{J_{A},L_{A}} \\
(-1)  ^{l+L_{F}+L_{I}}(4l+1)  (  2L_{A}+1)  C_{2l,\;J_{A},\;J_{F}}^{0,\;m_{F},\;m_{F}}C_{1,\;J_{A},\;J_{I}}^{\lambda,\;m_{F},\;m_{I}}
\pmqty{L_{A} & 2l & L_{F} \\ 0 & 0 & 0} \pmqty{L_{A} & 1 & L_{I} \\ 0 & 0 & 0}
\bmqty{J_{F} & 2l & J_{A} \\ L_{A} & S_{F} & L_{F}}  \bmqty{J_{I} & 1 & J_{A} \\ L_{A} & S_{I} & L_{I}}\\
\sum_{n_A,n_B}(  M_{B}-M_{A}) \pqty{\int\limits_{0}^{\infty}\dd{r}r^{2}(  R_{n_{F}L_{F}})^{*}j_{2l}\pqty{\frac{kr}{2}}  R_{n_{A}L_{A}}}
\pqty{\int\limits_{0}^{\infty}\dd{r}r^{2}(  R_{n_{A}L_{A}})^{*}r R_{n_{B}L_{I}}} \mel{J_{I},m_{I},n_{B}L_{I},s_{I}}{\mathcal{P}_{+}(\widehat{E}_{\alpha})}{\Psi_{I}} ,
\label{esp}
\end{multline}
where $M_{B}$ is the mass of the state $\ket{J_{I},m_{I},n_{B}L_{I},s_{I}}$, $M_{A}$ is the mass of $\ket{J_{A},m_{F},n_{A}L_{A},s_{F}}$, $C$ is the Clebsch-Gordan coefficient
\begin{equation}
C_{j_1,\;j_2,\;j_3}^{m_1,m_2,m_3}\equiv(-1)^{j_2-j_1-m_3}\sqrt{2 j_3+1}\pmqty{j_1 & j_2 & j_3 \\ m_1 & m_2 & -m_3}
\label{cg}
\end{equation}
with $(\;)  $ standing for the $3j$ symbol, and
\begin{equation}
\begin{bmatrix}
j_1 & j_2 & j_{12} \\
 j_3 & j & j_{23}
\end{bmatrix} \equiv
 (-1)^{j_1+j_2+j_3+j}\sqrt{(2j_{12}+1)(2j_{23}+1)}
\begin{Bmatrix}
j_1 & j_2 & j_{12} \\ 
j_3 & j & j_{23}
\end{Bmatrix}
\label{symbol}
\end{equation}
with $\{\;\}  $ standing for the $6j$ symbol.

As for the first magnetic term, we get from \eqref{tmec} and \eqref{OmagkFI}
\begin{multline}
(  \mathcal{M}_{J_{F},m_{F},J_{I},m_{I}}^{\lambda\;\text{(magnetic)}  _{\vb*{\sigma}\cp\vb*{k}}})  ^{\,{^{3}\!S_{1}}\rightarrow\gamma\,{^{3}\!P_{J}}} =   i \sqrt{2E_I}\sqrt{2E_F} e_c \lambda \frac{k}{m} \sum_{l=1}^\infty \sum_{\tilde J = \max(1,\abs{l-2})}^l \\
  i^{l}  \pqty{(-1)^l + (-1)^{S_F-S_I}} (-1)^{L_F+l-1} \frac{2l - 1}{2} \sqrt{3(2L_F+1)} C_{l-1,1,\tilde J}^{\;0,\;\lambda,\;\lambda}\,C_{\tilde J,\;J_F,\;J_I}^{\lambda,\; m_F,\; m_I} \pmqty{L_F & l-1 & L_I \\ 0 & 0 & 0} \\
\bmqty{S_I & 1 & S_F \\ \frac{1}{2} & \frac{1}{2} & \frac{1}{2}}  \bmqty{L_F & l-1 & L_I \\ S_F & 1 & S_I \\ J_F & \tilde J & J_I} \sum_{n_B} \pqty{\int\limits_{0}^{\infty}\dd{r}r^{2}(  R_{n_{F}L_{F}})^{*}j_{l-1}\pqty{\frac{kr}{2}}  R_{n_{B}L_{I}}} \mel{J_{I},m_{I},n_{B}L_{I},s_{I}}{\mathcal{K}(\widehat{E}_{\alpha})}{\Psi_{I}}.
\label{mspk}
\end{multline}

Finally, for the second magnetic term we obtain from \eqref{tmec} and \eqref{OmagpFI}
\begin{multline}
(  \mathcal{M}_{J_{F},m_{F},J_{I},m_{I}}^{\lambda\;\text{(magnetic)}  _{\vb*{\sigma}\cp\widehat{\vb*{p}}}})  ^{\,{^{3}\!S_{1}}\rightarrow
\gamma\,{^{3}\!P_{J}}} = i \sqrt{2E_I}\sqrt{2E_F} e_c \sqrt{\frac{3}{2}} \sum_{l=0}^\infty \sum_{J_A, L_A} i^l (-1)^{L_F+L_I}\pqty{(-1)^l+(-1)^{S_F-S_I}} (2 l +1) (2 L_A + 1) \\
C_{l,\;J_A,\; J_F}^{0,\;m_F,\;m_F} C_{1,\;J_A,\; J_I}^{\lambda,\;m_F,\;m_I}  \pmqty{L_A & l & L_F \\ 0 & 0 & 0} \pmqty{L_A & 1 & L_I \\ 0 & 0 & 0} \bmqty{J_F & l & J_A \\ L_A & S_F & L_F} \bmqty{S_I & 1 & S_F \\ \frac{1}{2} & \frac{1}{2} & \frac{1}{2}} \bmqty{L_A & 1 & L_I \\ S_F & 1 & S_I \\ J_A & 1 & J_I} \\
\sum_{n_A,n_B}(M_B - M_A)\pqty{\int\limits_{0}^{\infty}\dd{r}r^{2}(  R_{n_{F}L_{F}})^{*}j_{l}\pqty{\frac{kr}{2}}  R_{n_{A}L_{A}}}    \pqty{\int\limits_{0}^{\infty}\dd{r}r^{2}(  R_{n_{A}L_{A}})^{*}r R_{n_{B}L_{I}}} \mel{J_{I},m_{I},n_{B}L_{I},s_{I}}{\mathcal{P}_{-}(\widehat{E}_{\alpha})}{\Psi_{I}}.
\label{mspp}
\end{multline}
For simplicity the matrix elements of the energy dependent operators have been evaluated in momentum space.

\subsection{Transitions $^{3}\!P_{J}\rightarrow\gamma\,{^{3}\!S_{1}}$}

We proceed in the same manner for $^{3}\!P_{J}\rightarrow\gamma\,{^{3}\!S_{1}}$ transitions. Thus, from \eqref{tmec} and \eqref{OprimelFI} the electric part of the amplitude reads
\begin{multline}
(  \mathcal{M}_{J_{F},m_{F},J_{I},m_{I}}^{\lambda\;\text{\text{\text{(electric)}}}  })  ^{{^{3}\!P_{J}}\rightarrow\gamma\,{^{3}\!S_{1}}}= i\sqrt{2M_{I}}\sqrt{2E_{F}}\delta_{S_{I},S_{F}}e_{c} \sum_{l=0}^{\infty}\sum_{J_{A},L_{A}} \\
(-1)  ^{l}(4l+1)  \sqrt{(2 L_I + 1)(2 L_F + 1)}  C_{2l,\;J_{I},\;J_{A}}^{0,\;m_{I},\;m_{I}}C_{1,\;J_{F},\;J_{A}}^{\lambda,\;m_{F},\;m_{I}}
\pmqty{L_{I} & 2l & L_{A} \\ 0 & 0 & 0} \pmqty{L_{F} & 1 & L_{A} \\ 0 & 0 & 0}
\bmqty{J_{A} & 2l & J_{I} \\ L_{I} & S_{I} & L_{A}}  \bmqty{J_{A} & 1 & J_{F} \\ L_{F} & S_{F} & L_{A}}\\
\sum_{n_A,n_B}(  M_{A}-M_{F}) \pqty{\int\limits_{0}^{\infty}\dd{r}r^{2}(  R_{n_{F}L_{F}})^{*}r R_{n_{A}L_{A}}} \pqty{\int\limits_{0}^{\infty}\dd{r}r^{2}(  R_{n_{A}L_{A}})^{*}j_{2l}\pqty{\frac{kr}{2}}  R_{n_{B}L_{I}}} 
\mel{J_{I},m_{I},n_{B}L_{I},s_{I}}{\mathcal{P}_{+}(\widehat{E}_{\alpha})}{\Psi_{I}}.
\label{eps}
\end{multline}

where $M_{B}$ is the mass of the state $\ket{J_{I},m_{I},n_{B}L_{I},s_{I}}$ and $M_{A}$ is the mass of the intermediate state $\ket{J_{A},m_{I},n_{A}L_{A},s_{I}}$.

As for the first magnetic term, we get from \eqref{tmec} and \eqref{OprimemagkFI}
\begin{multline}
(  \mathcal{M}_{J_{F},m_{F},J_{I},m_{I}}^{\lambda\;\text{(magnetic)}  _{\vb*{\sigma}\cp\vb*{k}}})  ^{{^{3}\!P_{J}}\rightarrow\gamma\,{^{3}\!S_{1}}} =  i \sqrt{2E_I}\sqrt{2E_F} e_c \lambda \frac{k}{m} \sum_{l=1}^\infty \sum_{\tilde J = \max(1,\abs{l-2})}^l \\
  i^{l}  \pqty{(-1)^l + (-1)^{S_F-S_I}} (-1)^{L_F+l-1} \frac{2l - 1}{2} \sqrt{3(2L_F+1)}  C_{l-1,1,\tilde J}^{\;0,\;\lambda,\;\lambda}\,C_{\tilde J,\;J_F,\;J_I}^{\lambda,\; m_F,\; m_I} \pmqty{L_F & l-1 & L_I \\ 0 & 0 & 0} \\
 \bmqty{S_I & 1 & S_F \\ \frac{1}{2} & \frac{1}{2} & \frac{1}{2}}  \bmqty{L_F & l-1 & L_I \\ S_F & 1 & S_I \\ J_F & \tilde J & J_I} \sum_{n_B} \pqty{\int\limits_{0}^{\infty}\dd{r}r^{2}(  R_{n_{F}L_{F}})^{*}j_{l-1}\pqty{\frac{kr}{2}}  R_{n_{B}L_{I}}} \mel{J_{I},m_{I},n_{B}L_{I},s_{I}}{\mathcal{P}_{+}(\widehat{E}_{\alpha})}{\Psi_{I}}.
\label{mpsk}
\end{multline}

Finally, for the second magnetic term we obtain from \eqref{tmec} and \eqref{OprimemagpFI}
\begin{multline}
(  \mathcal{M}_{J_{F},m_{F},J_{I},m_{I}}^{\lambda\;\text{(magnetic)}  _{\vb*{\sigma}\cp\widehat{\vb*{p}}}})  ^{{^{3}\!P_{J}}\rightarrow
\gamma\,{^{3}\!S_{1}}} = i \sqrt{2E_I}\sqrt{2E_F} e_c \sqrt{\frac{3}{2}} \sum_{l=0}^\infty \sum_{J_A, L_A} i^l \pqty{(-1)^l+(-1)^{S_F-S_I}} (2 l +1) \sqrt{(2L_I+1)(2L_F+1)} \\
C_{l,\;J_I,\; J_A}^{0,\;m_I,\;m_I} C_{1,\;J_F,\; J_A}^{\lambda,\;m_F,\;m_I}  \pmqty{L_I & l & L_A \\ 0 & 0 & 0} \pmqty{L_F & 1 & L_A \\ 0 & 0 & 0} \bmqty{J_A & l & J_I \\ L_I & S_I & L_A} \bmqty{S_I & 1 & S_F \\ \frac{1}{2} & \frac{1}{2} & \frac{1}{2}} \bmqty{L_F & 1 & L_A \\ S_F & 1 & S_I \\ J_F & 1 & J_A} \\
\sum_{n_A,n_B}(M_A - M_F)\pqty{\int\limits_{0}^{\infty}\dd{r}r^{2}(  R_{n_{F}L_{F}})^{*}r R_{n_{A}L_{A}}}\pqty{\int\limits_{0}^{\infty}\dd{r}r^{2}(  R_{n_{A}L_{A}})^{*}j_{l}\pqty{\frac{kr}{2}}  R_{n_{B}L_{I}}} \mel{J_{I},m_{I},n_{B}L_{I},s_{I}}{\mathcal{P}_{-}(\widehat{E}_{\alpha})}{\Psi_{I}}.
\label{mpsp}
\end{multline}
\end{widetext}
For simplicity the matrix elements of the energy dependent operators have been evaluated in momentum space.

\section{Results and conclusions\label{SVI}}

From the electric and magnetic amplitudes detailed in the preceding section, the total amplitude
\begin{multline}
\mathcal{M}_{J_{F},m_{F},J_{I},m_{I}}^{\lambda}=\\
\mathcal{M}_{J_{F},m_{F},J_{I},m_{I}}^{\lambda\;\text{(electric)}  }+\mathcal{M}_{J_{F},m_{F},J_{I},m_{I}}^{\lambda\;\text{(magnetic)}_{\vb*{\sigma}\cp\vb*{k}}}+\mathcal{M}_{J_{F},m_{F},J_{I},m_{I}}^{\lambda\;\text{(magnetic)}  _{\vb*{\sigma}\cp\widehat{\vb*{p}}}}
\label{tme}
\end{multline}
and the width for the $I\rightarrow\gamma F$ decay, given by \eqref{width}, can be straightforwardly evaluated.

Notice that the intermediate $B$ states are $(  n_{B})  \,{^{3}\!S_{1}}$ for $^{3}\!S_{1}\rightarrow\gamma\,{^{3}\!P_{J}}$ and $(n_{B}){^{3}\!P_{J}}$ for $^{3}\!P_{J}\rightarrow\gamma\,{^{3}\!S_{1}}$ transitions. The intermediate $A$ states are $(  n_{A})  ^{3}P_{J_{A}}$ in both cases.

We have checked numerically that for the calculated $^{3}\!S_{1}\rightarrow\gamma\,{^{3}\!P_{J}}$ and ${^{3}\!P_{J}}\rightarrow\gamma\,{^{3}\!S_{1}}$ widths it is enough to take $n_{B}\leq2$ and $n_{A}\leq2$ to assure convergence at the level of $1\%$. As we did in the case of the $\frac{\vb*{p}_{c}}{M_{c}}$ approximation we have used the experimental masses for $2{^{3}\!P_{0}}$ (corresponding to $\chi_{c0}(  3860)$) and $2{^{3}\!P_{2}}$ (corresponding to $\chi_{c2}(  3930)$) and the Cornell calculated mass from Model~I for $2{^{3}\!P_{1}}$. The results obtained are shown in Table~\ref{tabcomplet}.

\begin{table*}
\begin{ruledtabular}
\begin{tabular}{lccc}
Radiative Decay & $(  \Gamma_\text{complete}^{(  \theor-\expt)  })  _{I}$ & $\Gamma_\expt^{PDG}$ & $(  \Gamma_\text{complete}^{(  \theor-\expt)  })  _{II}$ \\
& KeV & KeV & KeV \\
\hline
$\psi(  2S)  \rightarrow\gamma\chi_{c_{0}}(  1p)  $ &
$54$ & $28.8\pm1.4$ & $43$\\
$\psi(  2S)  \rightarrow\gamma\chi_{c_{1}}(  1p)  $ &
$35$ & $28.7\pm1.5$ & $30$\\
$\psi(  2S)  \rightarrow\gamma\chi_{c_{2}}(  1p)  $ &
$20$ & $28.0\pm1.3$ & $18$\\
$\chi_{c_{0}}(  1p)  \rightarrow\gamma J/\psi$ & $187$ & $151\pm14$
& $130$\\
$\chi_{c_{1}}(  1p)  \rightarrow\gamma J/\psi$ & $415$ & $288\pm22$
& $284$\\
$\chi_{c_{2}}(  1p)  \rightarrow\gamma J/\psi$ & $566$ & $374\pm27$
& $385$ \\
\end{tabular}
\end{ruledtabular}
\caption{Calculated widths as compared to data.for $\psi(2S)\rightarrow\gamma\chi_{c_{J}}(  1P)  $ and $\chi_{c_{J}}(1P)  \rightarrow\gamma J/\psi$ transitions. Notation as follows. $\Gamma_\text{complete}^{(  \theor-\expt)  }$: calculated width implemented with the experimental masses and photon energy. The subscripts $I$ and $II$ refer to Model~I and Model~II. $\Gamma_\expt^{PDG}$: measured widths \cite{PDG18}. }
\label{tabcomplet}
\end{table*}

A first general feature of these results is that the values of the widths from Model~I are systematically bigger than from Model~II. As this was also the case in the $\frac{\vb*{p}_{c}}{M_{c}}$ approximation, see Table~\ref{tabp/m}, it can be mostly attributed to the different wavefunctions from both models (let us recall that in the $\frac{\vb*{p}_{c}}{M_{c}}$ approximation the quark mass dependence of the calculated width is reduced with respect to the complete calculation).

A second general feature of the results in Table~\ref{tabcomplet} is that the calculated widths from Model~II are much closer to data than the ones from Model~I (the only exception being $2{^{3}\!S_{1}}\rightarrow\gamma\,1{^{3}\!P_{2}}$, where they are almost equal). More precisely, half (three) of the calculated widths from Model~II are within the experimental intervals; another one, the $1{^{3}\!P_{0}}\rightarrow\gamma\,1{^{3}\!S_{1}}$ calculated width is very close to the experimental value differing from it a $6\%,$ whereas the remaining two calculated widths, $2{^{3}\!S_{1}}\rightarrow\gamma\,1{^{3}\!P_{0,2}}$, differ from data about a $40\%$. In contrast, all the calculated widths from Model~I are out of the experimental intervals, in half of the cases with big differences  (bigger than $40\%$) with respect to data.

Centering on Model~II, the calculated widths show better agreement with data for ${2{^{3}\!S_{1}}\rightarrow \gamma\,1{^{3}\!P_{1}}}$ than for ${2{^{3}\!S_{1}}\rightarrow\gamma\,1{^{3}\!P_{0,2}}}$, and the same tendency, although attenuated, is observed for ${1{^{3}\!P_{1}}\rightarrow\gamma\,1{^{3}\!S_{1}}}$ as compared to $1{^{3}\!P_{0,2}}\rightarrow\gamma\,1{^{3}\!S_{1}}$. This points out to a good wavefunction for $1{^{3}\!P_{1}}$ and to some deficiency in the $1{^{3}\!P_{0,2}}$ wavefunctions, what can be correlated to the mass description of the charmonium states, see Table~\ref{tabIII}: quite good for $1{^{3}\!S_{1}}$, $2{^{3}\!S_{1}}$ and $1{^{3}\!P_{1}}$ and deficient for $1{^{3}\!P_{0,2}}.$ Then, it is quite possible that nonperturbative or second order perturbative corrections to the $1{^{3}\!P_{0,2}}$ wavefunctions play some role. If we assume that this is the case, so that that the wavefunction corrections are responsible for the observed discrepancies with data, then we may tentatively conclude that an accurate description of radiative decays in charmonium is feasible.

We may also conclude that the analysis of radiative decay processes serves as a stringent test of the spectroscopic quark model. In this regard, let us recall that the difference between Model~I and II comes essentially from the different values of the parameters, $\sigma$ and $M_{c}$. Actually, $\sigma$ and $M_{c}$ are correlated in the sense that an increase in
$\sigma$ has to be compensated with an increase in $M_{c}$ to maintain the spectral mass fit, see Table~\ref{tabIII}. The fact that Model~II works much better for the description of the studied radiative decays indicates that the value of the effective parameter $\sigma$ has to be quite close to its phenomenological value from the analysis of Regge trajectories in light mesons and, as a consequence, that the quark mass effective parameter, at least for the description of the low lying charmonium states, is constrained to be around $1.84$ GeV.

Certainly, the prediction of the decay widths for radiative transitions  not yet measured could provide, if confirmed by future experiments, strong support to our conclusions. In practice, this comparison program presents some difficulties. On the one hand we may expect that the $3{^{3}\!S_{1}}\rightarrow\gamma\,1{^{3}\!P_{0,2}}$ widths suffer at least from the same uncertainty as the $2{^{3}\!S_{1}}\rightarrow\gamma\,1{^{3}\!P_{0,2}}$. On the other hand the experimental resonance $\psi(4040)$ contains presumably a significant mixing of $3{^{3}\!S_{1}}$ and $2^{3}D_{1}$ states and the calculation of $2^{3}D_{1}\rightarrow\gamma\,1{^{3}\!P_{0,1,2}}$ beyond the $\frac{\vb*{p}_{c}}{M_{c}}$ approximation is quite uncertain due to the current dearth of $^{3}D_{J}$ state mass data to be implemented. The same arguments apply to $3{^{3}\!S_{1}}\rightarrow\gamma\,2{^{3}\!P_{0,2}}.$ On the other hand the observed decays $X(  3872)  \rightarrow \gamma\,(  1,2) {^{3}\!S_{1}}$ can not be described as $2 {^{3}\!P_1}  \rightarrow \gamma\,(  1,2) {^{3}\!S_{1}}$ from Model~II since the $X(3872)$ description requires the incorporation of meson-meson threshold effects. Furthermore, the same kind of effects might be present in $\chi_{c0}(3860)$ that we have assumed to be the pure $2{^{3}\!P_{0}}$ Cornell state in spite of being above its first $I(J^{PC})=0(0^{++})  $ meson-meson threshold, and in many other high lying charmonium sates.

This reduces the range of our comparison program to the prediction of the $2{^{3}\!P_{0,2}}\rightarrow \gamma \,(  1,2)\,{^{3}\!S_{1}}$ decay widths from Model~II with a significant degree of uncertainty due to the deficient description of the ${^{3}\!P_{0,2}}$ wavefunctions, and in the $2{^{3}\!P_{2}}\rightarrow \gamma \, 1\!  \,{^{3}\!S_{1}}$ case to the contribution of non diagonal matrix elements of the energy dependent operators. From our calculation, we may conservatively estimate a maximum uncertainty of about a $50\%$ in the results.

The results with the estimated maximum uncertainties are shown in Table~\ref{tabprediction}.

\begin{table}
\begin{ruledtabular}
\begin{tabular}{lc}
Radiative Decay & $(  \Gamma_\text{complete}^{(  \theor-\expt)  })  _{II}$ \\
& KeV \\
\hline
$\chi_{c2}(  3930)  \rightarrow\gamma J/\psi$ & $54 \pm 27$\\
$\chi_{c2}(  3930)  \rightarrow\gamma\psi(  2S)  $ & $128 \pm 64$\\
$\chi_{c0}(  3860)  \rightarrow\gamma J/\psi$ & $36 \pm 18$\\
$\chi_{c0}(  3860)  \rightarrow\gamma\psi(  2S)  $ & $45 \pm 23$\\
\end{tabular}
\end{ruledtabular}
\caption{Predicted widths from Model~II. Same notation as in Table~\ref{tabcomplet}. $\Gamma_\text{complete}^{(  \theor-\expt)  }$:
calculated width implemented with the experimental masses and photon energy. The errors correspond to the estimated maximum uncertainties.}
\label{tabprediction}
\end{table}

\section{Summary\label{SVII}}

In this article we have analyzed ${{^{3}\!S_1} \leftrightarrow {^{3}\!P_J}}$ electromagnetic transitions in charmonium. From two constituent quark models reasonably fitting the masses of the low lying charmonium states and from an elementary emission model for the description of the radiative processes we have shown that

\begin{enumerate}[i)]

\item neither the standard $\frac{\vb*{p}_{c}}{M_{c}}$ approximation to the electromagnetic transition operator nor its commonly used long wave length limit should be taken for granted for they can not give accurate account of data. Although this inaccuracy has been only quantitatively proved for two Cornell potential models it may be reasonably assumed to be valid in general since the Cornell potential form contains the more relevant terms for an appropriate description of the low lying charmonium states. As a consequence, the $\frac{\vb*{p}_{c}}{M_{c}}$ approximation or its long wave length limit should also not be used to discriminate between different spectroscopic quark models.

\item the use of the complete electromagnetic operator may allow for an accurate description of ${{^{3}\!S_1} \leftrightarrow {^{3}\!P_J}}$ radiative transitions between low lying charmonium states from a Cornell potential model, when the string tension parameter is close to the phenomenological value derived from the analysis of Regge trajectories in light mesons. This restricts drastically the possible values of the charm mass parameter to keep an appropriate charmonium mass description. Therefore, electromagnetic decay processes in charmonium can be used as a powerful tool to constrain the value of the charm mass parameter and to discriminate among different quark models.

\item the complete electromagnetic operator formalism developed for the calculation of the radiative widths can be straightforwardly applied to bottomonium $b\bar{b}$, and trivially extended to treat radiative decays of bottom, charmed mesons $b\bar{c}$ and $c\bar{b}$. Actually, for bottomonium the calculated widths are in agreement within a $15\%$ with the ones obtained in the $\frac{\vb*{p}_{b}}{M_{b}}$ approximation, the only exception being for ${3\,{^{3}\!S_1}\rightarrow \gamma \,1{^{3}\!P_J}}$ where the difference is about $50\%$. This confirms the sufficiency of the $\frac{\vb*{p}_{b}}{M_{b}}$  approach for calculating radiative decays in bottomonium, for the difference in the $3\,{^{3}\!S_1}\rightarrow \gamma 1\,{^{3}\!P_J}$ case is very uncertain due to the very relevant non diagonal contributions to the matrix element of the energy dependent operators.

\end{enumerate}

\begin{acknowledgments}
This work has been supported by \foreignlanguage{spanish}{Ministerio de Economía y Competitividad} of Spain (MINECO) and EU Feder Grant No.\ FPA2016-77177-C2-1-P, by SEV-2014-0398, and by EU Horizon 2020 research and innovation program under Grant No.\ 824093 (STRONG-2020). R.\,B.\ acknowledges a FPI fellowship from the \foreignlanguage{spanish}{Ministerio de Ciencia, Innovación y Universidades of Spain} under Grant No.\ BES-2017-079860.
\end{acknowledgments}

\appendix

\section{First quantized transition operator\label{firstop}}

Let us consider for instance the quark part of the vector component of the current (following the main text, we denote quark components by subscript $1$) given by

\begin{widetext}

\begin{multline}
\pqty{\vb*{j}(\vb*{x})}_1 = e_1 \left\{ -i \bqty{q_1^\dagger(\vb*{x}) \pqty{\frac{\grad{q_1(\vb*{x})}}{M_1 + {E}_1}}-\pqty{\frac{\grad{q_1^\dagger(\vb*{x})}}{M_1 +{E}_1}} q_1(\vb*{x})}\right.\\
\left.+\bqty{\pqty{\frac{\grad{q_1^\dagger(\vb*{x})}}{M_1 + {E}_1}} \cp \vb*{\sigma}\!_{1} \, q_1(\vb*{x})-q_1^\dagger(\vb*{x}) \vb*{\sigma}\!_{1} \cp \pqty{\frac{\grad{q_1(\vb*{x})}}{M_1 + {E}_1}}}\right\},
\end{multline}
and a quark Fock state given by
\begin{equation}
\left\vert \eta\right\rangle =\frac{1}{(  2\pi)  ^{\frac{3}{2}}}\int \dd{\vb{p}_1}\sum_{m_s}\phi_{\eta}^{m_s}(  \vb{p}_1)  b_{1}^{m_s\dagger}(  \vb{p}_1)  \ket{0} =\frac{1}{(  2\pi)  ^{\frac{3}{2}}}\int\dd{\vb{p}_1}\frac{1}{(  2\pi)  ^{\frac{3}{2}}}\int\dd{\vb{r}_1}e^{-i\vb{p}_1 \vdot\vb{r}_1} \sum_{m_s}\psi_{\eta}^{m_s}(  \vb{r}_1) b_{1}^{m_s\dagger}(  \vb{p}_1)\ket{0}
\end{equation}
where $\phi(  \vb{p}_1)  $ and $\psi(  \vb{r}_1)  $ stand for the wavefunction in momentum and configuration space respectively.

We define the first quantized quark part of the vector component of the current $(  \vb*{j}_\text{1st}(  \vb*{x}))_1$ by requiring that
\begin{equation}
\mel{\xi}{\pqty{\vb*{j}(  \vb*{x})}_1}{\eta} =\int \dd{\vb{r}_1}\sum_{m_s,m_s^\prime}\bigl(\psi_{\xi}^{m_s^\prime}(  \vb{r}_1)\bigr)^\ast (\vb*{j}_\text{1st}(  \vb*{x}))_1^{m_s^\prime, m_s} (\vb{r}_1) \psi_{\eta}^{m_s}(  \vb{r}_1) .
\end{equation}
It is then easy to check that
\begin{multline}
\pqty{\vb*{j}_\text{1st}(\vb*{x})}_1 = e_1 \sqrt{\frac{M_1 + {E}_1}{2 {E}_1}} \left[\delta^{(3)}(\vb*{x}-\vb*{r}_1) \frac{\vb*{p}_1}{M_1 + {E}_1} + \frac{\vb*{p}_1}{M_1 + {E}_1} \delta^{(3)}(\vb*{x}-\vb*{r}_1)\right.\\
\left.-i \vb*{\sigma}_1 \cp \pqty{\delta^{(3)}(\vb*{x}-\vb*{r}_1) \frac{\vb*{p}_1}{M_1 + {E}_1} - \frac{\vb*{p}_1}{M_1 + {E}_1} \delta^{(3)}(\vb*{x}-\vb*{r}_1)}\right]\sqrt{\frac{M_1 + {E}_1}{2 {E}_1}} .
\label{jvecQ}
\end{multline}
By using the radiation gauge so that the time component of the electromagnetic field vanishes $(  A^{0}(  \vb*{x})=0)  $ we get
\begin{multline}
\mathcal{H}^{int} = \int \dd{\vb*{x}}j^{\mu}_\text{1st}(\vb*{x})A_{\mu}(  \vb*{x}) = - \int \dd{\vb*{x}}\vb*{j}_\text{1st}(\vb*{x})\vdot \vb*{A}(  \vb*{x})\\
 =-\sum_{\alpha=1,2} e_\alpha \sqrt{\frac{M_\alpha +{E}_\alpha}{2 {E}_\alpha}} \left[\pqty{\vb*{A}(\vb*{r}_\alpha) \vdot \frac{\vb*{p}_\alpha}{M_\alpha + {E}_\alpha} + \frac{\vb*{p}_\alpha}{M_\alpha + {E}_\alpha} \vdot \vb*{A}(\vb*{r}_\alpha)}\right.\\
\left.-i \pqty{ \vb*{\sigma}_{\alpha} \cp \vb*{A}(\vb*{r}_\alpha) \vdot \frac{\vb*{p}_\alpha}{M_\alpha + {E}_\alpha} - \frac{\vb*{p}_\alpha}{M_\alpha + {E}_\alpha} \vdot \vb*{\sigma}_{\alpha} \cp \vb*{A}(\vb*{r}_\alpha) }\right]\sqrt{\frac{M_\alpha + {E}_\alpha}{2 {E}_\alpha}} .
\end{multline}
To obtain the desired form of the transition operator we expand $\vb*{A}(\vb*{x})$, the electromagnetic field at $t=0$, in terms of creation and annihilation operators (see for example \cite[p.~123]{Pes95}):
\begin{equation}
\vb*{A}(\vb*{x}) = \int\frac{\dd{\vb{k}}}{(2\pi)^3}\frac{1}{\sqrt{2 \mathrm{k}_0}} \sum_{\lambda=\pm1} \pqty{\vb*{\epsilon}_{\vb{k}}^{\lambda} \, a_{\vb{k}}^\lambda e^{i \vb{k}\vdot\vb*{x}}+(  \vb*{\epsilon}_{\vb{k}}^{\lambda})^{*} \, a_{\vb{k}}^{\lambda\dagger} e^{-i \vb{k}\vdot\vb*{x}}}
\label{elmanop}
\end{equation}
where $a_{\vb{k}}^{\lambda\dagger}$  is the operator that creates a photon with three-momentum $\vb{k}$ and polarization $\lambda$
\begin{equation}
\ket{\vb{k},\lambda} = a_{\vb{k}}^{\lambda\dagger} \ket{0} .
\end{equation}
Then, taking the matrix element of \eqref{elmanop} between the vacuum and the one photon state yields
\begin{equation}
\mel{\vb*{k},\lambda}{\vb*{A}(  \vb*{x})}{0} =\frac{1}{\sqrt{2k_{0}}}e^{-i\vb*{k}\vdot\vb*{x}}(  \vb*{\epsilon}_{\vb*{k}}^{\lambda})^{*}
\label{kame}
\end{equation}
where $\vb*{\epsilon}_{\vb*{k}}^{\lambda}$ stands for the photon polarization vector, and the first quantized transition operator to be sandwiched between the meson states reads
\begin{multline}
\mel{\vb*{k},\lambda}{\mathcal{H}^{int}_\text{1st}}{0} = -\frac{1}{\sqrt{2 k_0}} \sum_{\alpha=1,2} e_\alpha \sqrt{\frac{M_\alpha + \widehat{E}_\alpha}{2 \widehat{E}_\alpha}} \left[e^{-i\vb*{k}\vdot\widehat{\vb*{r}}_\alpha}\frac{\widehat{\vb*{p}}_\alpha}{M_\alpha+\widehat{E}_\alpha} + \frac{\widehat{\vb*{p}}_\alpha}{M_\alpha+\widehat{E}_\alpha}e^{-i\vb*{k}\vdot\widehat{\vb*{r}}_\alpha}\right. \\
\left. -i \vb*{\sigma}\!_{\alpha} \cp \pqty{e^{-i\vb*{k}\vdot\widehat{\vb*{r}}_\alpha} \frac{\widehat{\vb*{p}}_\alpha}{M_\alpha+\widehat{E}_\alpha} - \frac{\widehat{\vb*{p}}_\alpha}{M_\alpha+\widehat{E}_\alpha} e^{-i\vb*{k}\vdot\widehat{\vb*{r}}_\alpha}}\right] \vdot (  \vb*{\epsilon}_{\vb*{k}}^{\lambda})^{*} \sqrt{\frac{M_\alpha + \widehat{E}_\alpha}{2 \widehat{E}_\alpha}} .
\end{multline}

\section{Gauge invariance\label{gaugeinv}}

Let us show that the electromagnetic current satisfies the condition
\begin{equation}
k_{\mu}j^{\mu}_\text{1st}(\vb*{x})=0,
\end{equation}
meaning that the current is conserved as required by gauge invariance. For this purpose, we have to verify that the matrix element of $k_{\mu}j^{\mu}_\text{1st}(\vb*{x})$ between any initial and final meson state is zero. Equivalently, expanding the meson states in terms of quark and antiquark momentum eigenstates, we can verify that
\begin{equation}
\begin{split}
\mel{\vb*{p}_{f},m'_{s}}{k_{\mu}j^{\mu}_\text{1st}(\vb*{x})}{\vb*{p}_{i},m_s} &= \mel{\vb*{p}_{f},m'_{s}}{k_{0}j^{0}_\text{1st}(\vb*{x})-\vb*{k}\vdot \vb*{j}_\text{1st}(\vb*{x})}{\vb*{p}_{i},m_s} \\
&=k_{0}\mel{\vb*{p}_{f},m'_{s}}{ j^{0}_\text{1st}(\vb*{x})}{\vb*{p}_{i},m_s} -\vb*{k}\vdot\mel{\vb*{p}_{f},m'_{s}}{\vb*{j}_\text{1st}(\vb*{x})}{\vb*{p}_{i},m_s} =0
\end{split}
\end{equation}
where $\vb*{p}_{i}$ is the three-momentum of the initial quark state going to a final state with a photon and a quark with three-momentum $\vb*{p}_{f}$.

Let us prove it for the quark contribution. The time component of the current is given by
\begin{multline}
\pqty{j_\text{1st}^{0}(  \vb*{x})}_1 = e_1 \sqrt{\frac{M_1 +{E}_1}{2 {E}_1}} \Biggr[\delta^{(3)}(\vb*{x}-\vb*{r}_1)\\
+\left(\frac{\vb*{p}_1}{M_1 +{E}_1}\vdot \delta^{(3)}(\vb*{x}-\vb*{r}_1) \frac{\vb*{p}_1}{M_1 + {E}_1}+ i \vb*{\sigma}_1\vdot \frac{\vb*{p}_1}{M_1 + {E}_1} \cp \delta^{(3)}(\vb*{x}-\vb*{r}_1) \frac{\vb*{p}_1}{M_1 +{E}_1}\right)\Biggr] \sqrt{\frac{M_1 + {E}_1}{2 {E}_1}}
\label{j0Q}
\end{multline}
as can be easily checked following the procedure detailed in Appendix~\ref{firstop}, whilst the vector component is given by \eqref{jvecQ}.

To calculate the matrix elements we take into account the action of the energy and momentum operators:
\begin{align}
\vb*{p}_1\ket{\vb*{p}_{i}, m_s} &= \vb*{p}_{i} \ket{\vb*{p}_{i}, m_s},\\
\bra{\vb*{p}_{f}, m'_s}\vb*{p}_1 &= \bra{\vb*{p}_{f}, m'_s}\vb*{p}_{f},
\end{align}
and
\begin{align}
{E}_1 \ket{\vb*{p}_{i}, m_s} &= \sqrt{M_1^2+\vb*{p}_{i}^2} \ket{\vb*{p}_{i},m_s} = E_{i} \ket{\vb*{p}_{i},m_s},\\
\bra{\vb*{p}_{f}, m'_s} {E}_1 &= \bra{\vb*{p}_{f}, m'_s} \sqrt{M_1^2+\vb*{p}_{f}^2} = \bra{\vb*{p}_{f}, m'_s} E_{f}.
\end{align}
The matrix element of the time component reads
\begin{multline}
\mel{\vb*{p}_f,m'_s}{\pqty{j_\text{1st}^0(\vb*{x})}_1}{\vb*{p}_i,m_s} = e_1 \Biggr[\delta^{m'_s,m_s}\pqty{1+ \frac{\vb*{p}_i}{M_1 + E_i} \cdot \frac{\vb*{p}_f}{M_1 + E_f}} - i \vb*{\sigma}_1^{m'_s,m_s}\vdot \frac{\vb*{p}_i}{M_1 + E_i} \cp \frac{\vb*{p}_f}{M_1 + E_f}\Biggr] \\
\sqrt{\frac{M_1 + E_f}{2 E_f}} \sqrt{\frac{M_1 +E_i}{2E_i}} e^{i (\vb*{p}_i - \vb*{p}_f)\vdot \vb*{x}}
\end{multline}
where we have used
\begin{equation}
\mel{\vb*{p}_{f},m'_{s}}{\delta(  \vb*{x}-\vb{r}_1)}{\vb*{p}_{i},m_s} = e^{i (\vb*{p}_i - \vb*{p}_f)\vdot \vb*{x}} \delta^{m'_s,m_s}.
\end{equation}
Note that for momentum eigenstates we have adopted the nonrelativistic normalization
\begin{equation}
\braket{\vb*{p}^\prime,m'_s}{\vb*{p},m_s}=(2\pi)^3 \delta^{(3)}(\vb*{p}-\vb*{p}^\prime)\delta^{m^\prime_s,m_s},
\end{equation}
consistently with \cite[p.~567]{PDG18}. In the same way, for the matrix element of the vector component we obtain
\begin{multline}
\mel{\vb*{p}_f,m'_s}{\pqty{\vb*{j}_\text{1st}(\vb*{x})}_1}{\vb*{p}_i,m_s} = e_1  \left[\delta^{m'_s,m_s}\pqty{\frac{\vb*{p}_i}{M_1 + E_i} + \frac{\vb*{p}_f}{M_1 +E_f}} -i \vb*{\sigma}_1^{m'_s,m_s} \cp \pqty{\frac{\vb*{p}_i}{M_1 + E_i} - \frac{\vb*{p}_f}{M_1 + E_f} }\right] \\
\sqrt{\frac{M_1 + E_f}{2 E_f}} \sqrt{\frac{M_1 +E_i}{2E_i}} e^{i (\vb*{p}_i - \vb*{p}_f)\vdot \vb*{x}} .
\end{multline}
\end{widetext}

Now, using
\begin{equation}
\vb*{p}_{i,f} \vdot \vb*{p}_{i,f} = \vb*{p}_{i,f}^2 = E_{i,f}^2 - M_1^2
\end{equation}
it is easy to show that
\begin{multline}
 (\vb*{p}_{i}-\vb*{p}_{f}) \vdot \pqty{\frac{\vb*{p}_i}{M_1 + E_i} + \frac{\vb*{p}_f}{M_1 +E_f}}\\
 = (E_i - E_f) \pqty{1+\frac{\vb*{p}_i}{M_1 + E_i} \vdot \frac{\vb*{p}_f}{M_1 + E_f}}
\end{multline}
and
\begin{multline}
(\vb*{p}_{i}-\vb*{p}_{f}) \vdot \vb*{\sigma}_1^{m'_s,m_s} \cp \pqty{\frac{\vb*{p}_i}{M_1 + E_i} - \frac{\vb*{p}_f}{M_1 + E_f} } \\= (E_i - E_f) \vb*{\sigma}_1^{m'_s,m_s} \vdot \frac{\vb*{p}_i}{M_1 + E_i} \cp \frac{\vb*{p}_f}{M_1 + E_f} ,
\end{multline}
so that
\begin{multline}
(\vb*{p}_{i}-\vb*{p}_{f}) \vdot \mel{\vb*{p}_f,m'_s}{\pqty{\vb*{j}_\text{1st}(\vb*{x})}_1}{\vb*{p}_i,m_s} \\= (E_{i}-E_{f}) \mel{\vb*{p}_f,m'_s}{\pqty{j^0_\text{1st}(\vb*{x})}_1}{\vb*{p}_i,m_s}.
\label{easyeq}
\end{multline}

Then, applying four-momentum conservation in the emission of the photon by the quark
\begin{align}
k_{0}&=E_{i}-E_{f} , \\
\vb*{k} &= \vb*{p}_{i}-\vb*{p}_{f} ,
\end{align}
\eqref{easyeq} becomes
\begin{multline}
\vb*{k} \vdot \mel{\vb*{p}_f,m'_s}{\pqty{\vb*{j}_\text{1st}(\vb*{x})}_1}{\vb*{p}_i,m_s}  \\= k_0 \mel{\vb*{p}_f,m'_s}{\pqty{j^0_\text{1st}(\vb*{x})}_1}{\vb*{p}_i,m_s} ,
\end{multline}
showing that the quark contribution to the current is conserved.

It is clear that the substitutions $e_1\rightarrow e_2$ and $M_1\rightarrow M_2$ do not affect this result, so that the the antiquark contribution is conserved as well. Then, the conservation of both the quark and antiquark components separately implies the conservation of the total current defined by \eqref{cur0} and \eqref{curvec}.

\bibliography{emcharmbib}

\end{document}